\documentclass{IEEEtran}
\usepackage[latin9]{inputenc}
\usepackage{color}
\usepackage{array}
\usepackage{rotfloat}
\usepackage{mathrsfs}
\usepackage{amsmath}
\usepackage{amssymb}
\usepackage{graphicx}

\makeatletter

\providecommand{\tabularnewline}{\\}

\usepackage{cite}
\usepackage{amsfonts}\usepackage{algorithmic}
\def\BibTeX{{\rm B\kern-.05em{\sc i\kern-.025em b}\kern-.08em
    T\kern-.1667em\lower.7ex\hbox{E}\kern-.125emX}}

\makeatother

\begin{document}

\title{Tapir: Automation Support of Exploratory Testing Using Model Reconstruction
of the System Under Test }

\author{Miroslav Bures, Karel Frajtak, and Bestoun S. Ahmed \thanks{M. Bures, Software Testing Intelligent Lab (STILL), Department of
Computer Science, Faculty of Electrical Engineering Czech Technical
University, Karlovo nam. 13, 121 35 Praha 2, Czech Republic, (email:
buresm3@fel.cvut.cz)} \thanks{K. Frajtak, Software Testing Intelligent Lab (STILL), Department of
Computer Science, Faculty of Electrical Engineering Czech Technical
University, Karlovo nam. 13, 121 35 Praha 2, Czech Republic, (email:
frajtak@fel.cvut.cz)} \thanks{B. Ahmed, Software Testing Intelligent Lab (STILL), Department of
Computer Science, Faculty of Electrical Engineering Czech Technical
University, Karlovo nam. 13, 121 35 Praha 2, Czech Republic, (email:
albeybes@fel.cvut.cz)}}
\maketitle
\begin{abstract}
For a considerable number of software projects, the creation of effective
test cases is hindered by design documentation that is either lacking,
incomplete or obsolete. The exploratory testing approach can serve
as a sound method in such situations. However, the efficiency of this
testing approach strongly depends on the method, the documentation
of explored parts of a system, the organization and distribution of
work among individual testers on a team, and the minimization of potential
(very probable) duplicities in performed tests. In this paper, we
present a framework for replacing and automating a portion of these
tasks. A screen-flow-based model of the tested system is incrementally
reconstructed during the exploratory testing process by tracking testers\textquoteright{}
activities. With additional metadata, the model serves for an automated
navigation process for a tester. Compared with the exploratory testing
approach, which is manually performed in two case studies, the proposed
framework allows the testers to explore a greater extent of the tested
system and enables greater detection of the defects present in the
system. The results show that the time efficiency of the testing process
improved with framework support. This efficiency can be increased
by team-based navigational strategies that are implemented within
the proposed framework, which is documented by another case study
presented in this paper.
\end{abstract}


\section{Introduction}

\IEEEPARstart{T}{he} contemporary software development market is
characterized by the increasing complexity of implemented systems,
a decrease in the time to market, and a demand for real-time operation
of these systems on various mobile devices \cite{Christof2017}. An
adequate software system is needed to solve the quality-related problems
that arise from this situation. Model-Based Testing (MBT) is a suitable
method for generating efficient test cases \cite{Lindvall2016}. For
a considerable ratio of the cases, the direct applicability of the
method is hindered by the limited availability and consistency of
the test basis, which is used to create the model.

To overcome these limitations, we explore possible crossover between
common MBT techniques and the exploratory testing approach. Exploratory
testing is defined as the simultaneous testing, learning, documentation
of the System Under Test (SUT) and creation of the test cases \cite{pfahl2014exploratory}.
The exploratory testing approach is a logical choice for testing systems
for which a suitable test basis is not available. Even when the test
basis is available, and the test cases are created, they can be either
obsolete or inconsistent and structured at an excessively high level
\cite{Gebizli2017}. Thus, testers employ the exploratory testing
technique as a solution for overcoming these obstacles. 

The key factors for the efficiency of exploratory testing are consistent
documentation of the explored path and exercised test cases \cite{Raappana2016,pfahl2014exploratory}.
This systematic documentation has the following features:
\begin{enumerate}
\item prevents the duplication of (informal) test scenarios that are executed
by various testers, which prevents a waste of resources;
\item leads to an exploration of the parts of the SUT that have not been
previously explored and tested;
\item improves the efficiency and accuracy of the defect reporting process;
and
\item improves the transparency and documentation of the testing process,
which is necessary for reporting and making managerial decisions related
to a project.
\end{enumerate}
Regarding the weaknesses of the exploratory testing, several issues
have been observed in previous reports. Particularly, the low level
of structuring of the testing process and a certain \textit{ad hoc}
factor can prevent efficient management of the exploratory testing
process. These issues may cause problems with the prioritization of
the tests, the selection of suitable tests in the actual state of
the testing process and the repetition of the exercised tests \cite{pfahl2014exploratory,shah2014towards}. 

Particular SUT exploration strategies are considered important in
the exploratory testing process \cite{Micallef2016}. To efficiently
conduct an exploration strategy, the exercised tests must be manually
recorded and documented, which can generate additional overhead for
the testing team. This overhead can outweigh the benefits gained by
a more efficient SUT exploration strategy. Thus, an automation of
the tasks related to the documentation of the exercised tests is an
option worth exploring.

Teamwork can have a significant role in exploratory testing. The team
organization increases the efficiency of the exploratory testing technique
in terms of identified defects \cite{Raappana2016}, and it may also
prevent repetitive tests and test data combinations. However, these
possibilities have only been explored in a manual version of the exploratory
testing process \cite{Raappana2016}.

These issues represent our motivation for exploring the possibilities
of supporting exploratory testing by a suitable machine support method.

This paper summarizes a concept of such a machine support. The paper
focuses on SUT exploration strategies and the generation of high-level
test cases from the SUT model. The model is reengineered during the
exploratory testing activities. The paper presents the Test Analysis
SUT Process Information Reengineering (\textbf{Tapir}) framework,
which guides exploratory testers during their exploration of a SUT.
The first objective of the framework is to enable the testing team
members to explore the SUT more systematically and efficiently than
the manual approach. The second objective of the framework is to automatically
document the explored parts of the SUT and create high-level test
cases, which guide the testers in the SUT. In this process, the framework
continuously builds a SUT model front-end UI. This model is enriched
by numerous elements that enhance the testing process.

This paper is organized as follows. Section \ref{sec:Real-time-construction-of}
presents the principle of the Tapir framework and introduces the SUT
model and its real-time construction process. Section \ref{sec:Generation-of-navigational}
explains the process of guiding the exploratory tester in the SUT
based on this model. Section \ref{sec:Experimental-Setup} describes
the setup of the performed case studies. Section \ref{sec:Experiment-Results-and}
presents the results of these case studies. Section \ref{subsec:Evaluation-of-the}
discusses the presented results. Section \ref{sec:Threats-to-validity}
discusses threats to the validity of the results, and Section \ref{sec:Related-work}
presents related studies. Finally, Section \ref{sec:Conclusion} provides
the concluding remarks of the paper.

\section{\label{sec:Real-time-construction-of}Real-time Construction of the
SUT Model}

\textcolor{black}{This section summarizes the functionality of the
Tapir framework and the underlying SUT model.}

\subsection{\label{subsec:Principle-of-the}Principles of the Tapir Framework }

The aim of the Tapir framework is to improve the efficiency of exploratory
testing by automating the activities related to the following:
\begin{enumerate}
\item records of previous test actions in the SUT,
\item decisions regarding the parts of the SUT that \textcolor{red}{will
be explored} in the subsequent test steps, and
\item organization of work for a group of exploratory testers.
\end{enumerate}
The framework tracks a tester\textquoteright s activity in the browser
and incrementally builds the SUT model based on its \textcolor{black}{User
Interface (UI)}. Based on this model, which can be extended by the
tester\textquoteright s inputs, navigational test cases are generated.
The explored paths in the SUT are recorded for the individual exploratory
testers. The navigational test cases help the testers explore the
SUT more efficiently and systematically, especially when considering
the teamwork of a more extensive testing group (typically larger than
five testers).

\begin{figure*}
\begin{centering}
\includegraphics[scale=0.45]{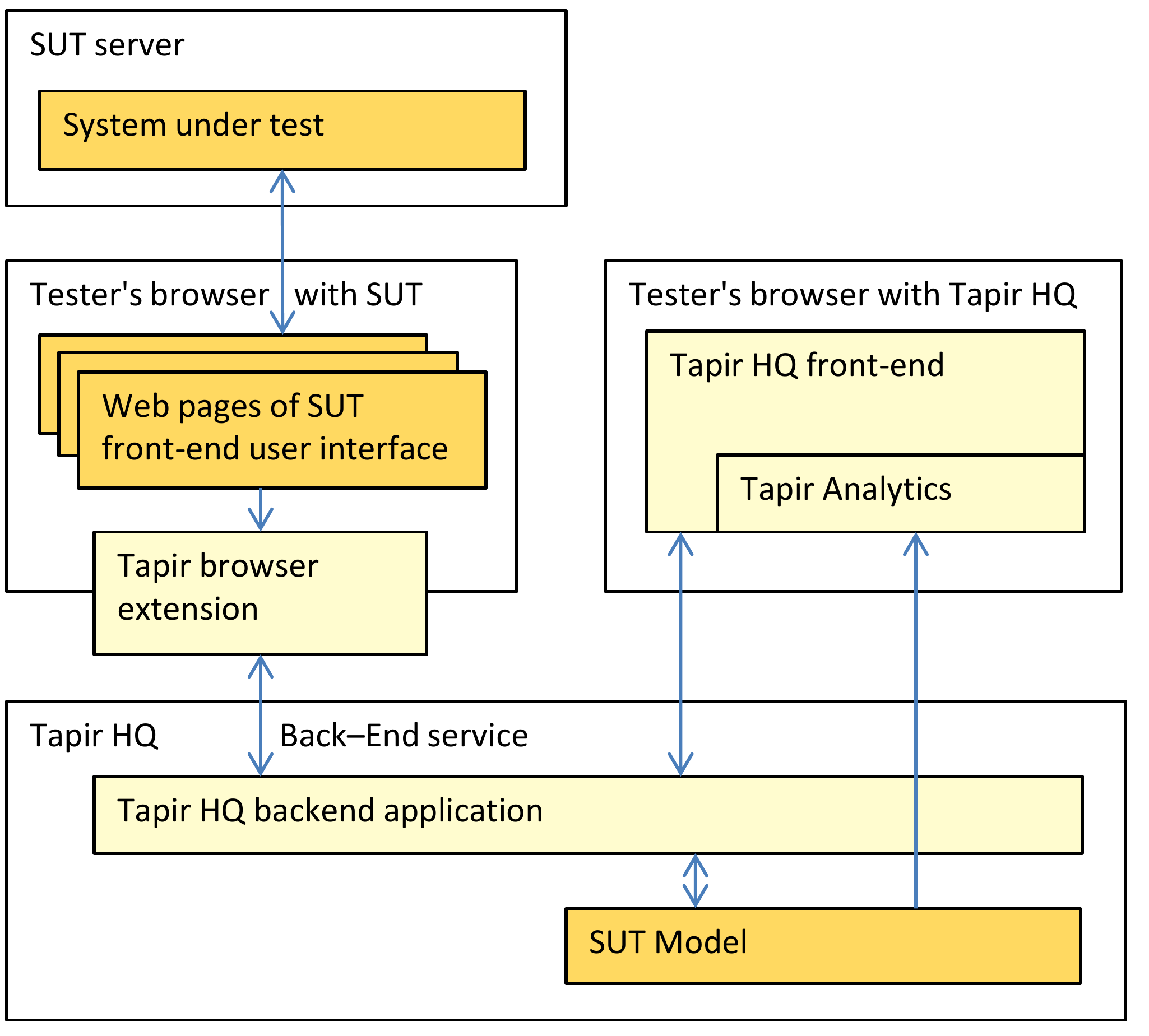}
\par\end{centering}
\caption{Overall architecture of the Tapir framework\label{fig:Overal-architecture-of}}
\end{figure*}

Technically, the Tapir framework consists of three principle parts.
\begin{enumerate}
\item \textbf{Tapir Browser Extension}: this extension tracks a tester\textquoteright s
activity in the SUT and sends the required information to the Tapir
HQ component, and it also highlights the UI elements of the SUT in
a selected \textcolor{black}{mode (i.e., the elements already analyzed
by the Tapir framework or elements suggested for exploration in the
tester\textquoteright s next step). The extension also analyzes the
SUT pages during the building of the SUT model. Currently, implemented
for the Chrome browser.}
\item \textbf{Tapir HQ}: this part is implemented as a standalone web application
that guides the tester through the SUT, provides navigational test
cases, and enables a Test Lead to prioritize the pages and links and
enter suggested Equivalence Classes (ECs) for the SUT inputs and related
functionality. This part constructs and maintains the SUT model. Tapir
HQ runs in a separate browser window or tab, thus serving as a test
management tool that displays the test cases for the SUT.
\item \textbf{Tapir Analytics}: this component enables the visualization
of the current state of the SUT model and a particular state of SUT
exploration. This part is also implemented as a module of Tapir HQ
that shares the SUT model with the Tapir HQ application.
\end{enumerate}
The overall system architecture is depicted in Figure \ref{fig:Overal-architecture-of}.

The Tapir framework defines two principal user roles.
\begin{enumerate}
\item \textbf{Tester}: a team member who is guided by the Tapir framework
to explore the parts of the SUT that have not been previously explored.
For each of the testers, a set of navigational strategies and a test
data strategy can be set by the Test Lead. The navigational strategy
determines a sequence of the SUT functions to be explored during the
tests, which is suggested to the tester by the navigational test cases.
The test data strategy determines the test data to enter on the SUT
forms and similar inputs during the tests. The test data are also
suggested to the testers by the navigational test cases.
\item \textbf{Test Lead}: senior team member who explores the SUT before
asking the testers to perform detailed tests. In addition to the tester\textquoteright s
functionalities, the Test Lead has the following principal functionalities. 
\begin{enumerate}
\item Prioritization of the pages, links, and UI action elements of the
SUT. During the first exploration, the Test Lead can determine the
priority of the particular screens and related elements. This priority
is saved as a part of the SUT model and subsequently employed in the
navigational strategies (refer to Section \ref{subsec:Navigational-strategies}). 
\item Definition of suitable input test data. During the first exploration,
the Test Lead can define ECs for the individual input fields that
are detected on a particular page. The ECs are saved to the SUT model
and subsequently employed in the process when generating navigational
test cases. After the ECs are defined for all inputs of the form on
the particular page, the Test Lead can let the Tapir framework generate
the test data combinations using an external Combinational Interaction
Testing (CIT) module. This module generates a smaller number of efficient
test data combinations based on a set of inputs on the page and the
ECs defined for these inputs. In this process, a Particle Swarm Test
Generator (PSTG) algorithm \cite{ahmed2012application}\textcolor{red}{{}
is applied}. Additional details are provided in Section \ref{subsec:The-Tapir-architecture}.
\end{enumerate}
\end{enumerate}
The Test Lead can dynamically change the navigational and test data
strategy of particular testers during the testing to reflect the current
state and priorities in the testing process. These strategies are
explained in Sections \ref{subsec:Navigational-strategies} and \ref{subsec:Test-data-strategies}.
The role of the Test Lead is not mandatory in the process. The Tapir
framework can be employed by a team of exploratory testers without
defining this role. In this case, functions related to prioritization
and test data definitions are not available to the team. The framework
administrator sets navigational strategies for team members.

\subsection{\label{subsec:The-SUT-Model}SUT Model}

For the purpose of systematic navigation by exploratory testers in
the SUT, we evolved the following model during our work and experiments
with the Tapir framework. $T$ denotes all exploratory testers who
are testing the SUT. The set $T$ includes testers $t_{1}...t_{n}$.
A tester can be given the role of Test Lead. The SUT is defined as
a tuple $\left(W,I,A,L\right)$, where $\mathit{W}$ is a set of SUT
pages, $\mathit{I}$ is a set of input elements displayed to the user
on the web pages of the SUT user interface, $\mathit{A}$ is a set
of action elements (typically \texttt{\textless{}form\textgreater{}}
element submits), and $\mathit{L}$ is a set of link elements displayed
to the user. 

A Web Page $w\in W$ is a tuple  $\left(I_{w},A_{w},L_{w},\Theta_{w},\phi_{w},M_{w}\right)$,
where $I_{w}\subseteq I$ is a set of input elements, $A_{w}\subseteq A$
is a set of action elements and $L_{w}\subseteq L$ is a set of link
elements located on page $\mathit{w}$. A Web Page $w$ can contain
action elements that can perform actions with more than one form displayed
on the page. In our notation, $I_{a}\subseteq I_{w}$ contains a set
of input elements that are connected to the action elements $a\in A_{w}$. 

On the SUT page $w$, an input element is a data entry field (text
field, drop-down item, checkbox or another type of input element defined
by the HTML for the forms). An action element triggers the submission
of the form, invokes data processing in the SUT and transition to
the next page $w_{next}$. The action elements are HTML buttons or
elements invoking submit event on the page. The link elements are
HTML links, or elements that invoke a transition to the next page
$w_{next}$. Typically, the SUT header or footer menu is captured
by the link elements.

Then, $range(i)$ denotes the particular data range that can be entered
in an input element $i\in I_{w}$. The $range(i)$ can be either an
interval or a set of discrete values (items of a list of values for
instance). Then, $range(I_{w})$ contains these ranges for the input
elements of $I_{w}$.

$\Theta_{w}$ is a set of action transition rules $\theta:w\times range(I_{w})\times A_{w}\rightarrow w_{next}$,
where $\mathit{w_{next}}\in W$ is a SUT web page that is displayed
to a user as a result of exercising the action element $a\in A_{w}$
with particular input data entered in the input elements $I_{a}\subseteq I_{w}$. 

$\varPhi_{w}$ is a set of action transition rules $\phi:w\times L{}_{w}\rightarrow w_{next}$,
where $\mathit{w_{next}}\in W$ is a SUT web page that is displayed
to the user as a result of exercising a link element $l\in L_{w}$.

Web Pages that are accessible from \textcolor{red}{Web Page $w$}
by the defined transition rules $\Theta_{w}$ and $\varPhi_{w}$ are
denoted as $next(w)$.

$\mathit{M}\subseteq W$ is a set of UI master pages. The Master Page
models repetitive components of the SUT user interface, such as a
page header with a menu or a page footer with a set of links. The
definition of the Master Page is the same as the definition of a Web
Page, and the Master Pages can be nested (refer to the Web Page definition).
$M_{w}\in M$ is a set of Master Pages of page $w$, and this set
can be empty. Additionally, $w_{0}\in W$ represents the home page
(defined page, from which the exploratory testers start exploring
the SUT) and $w_{e}\in W$ represents the standard error page displayed
during fatal system malfunctions (e.g., the exception page in J2EE
applications).

\begin{figure*}
\begin{centering}
\includegraphics[scale=0.5]{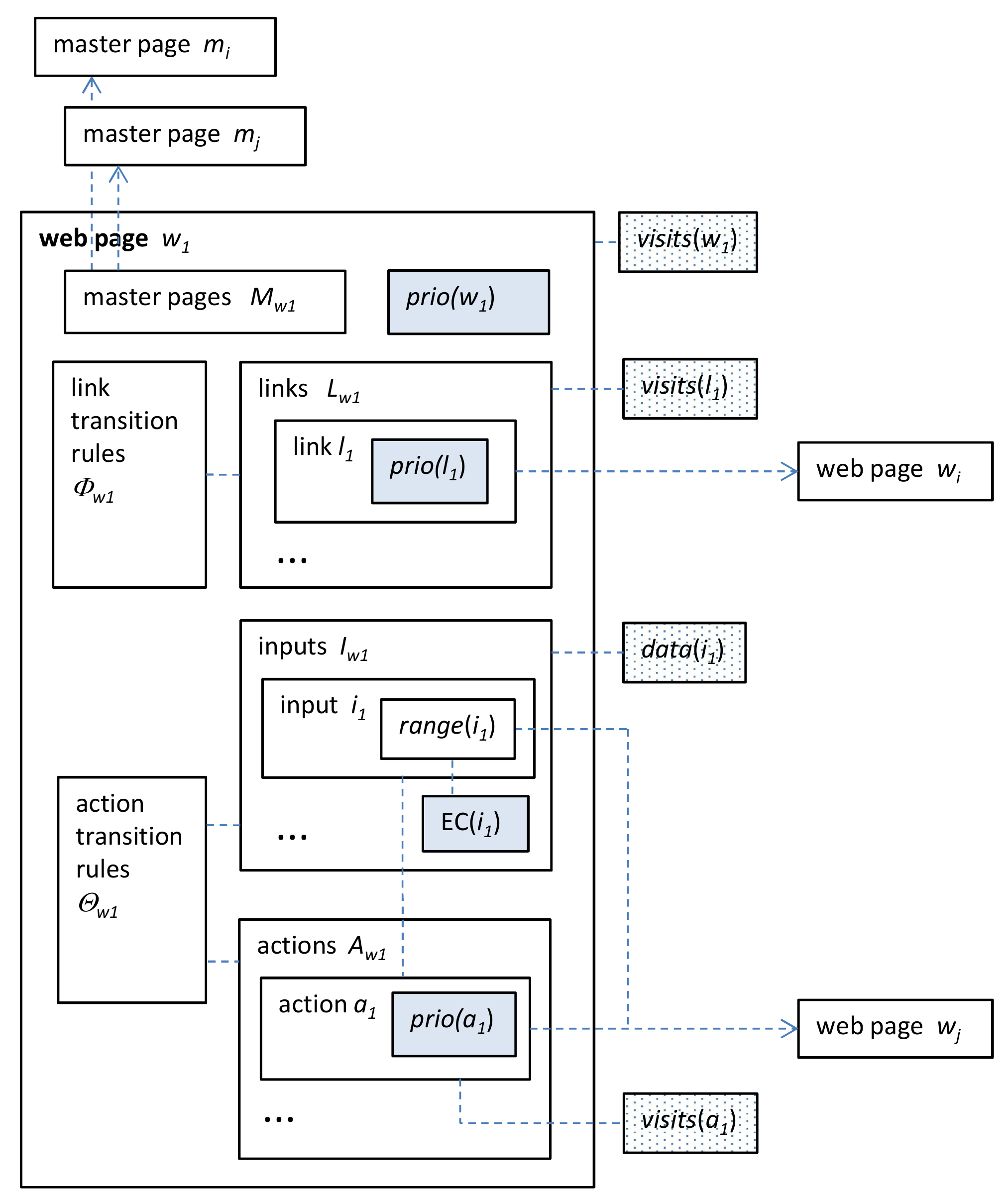}
\par\end{centering}
\caption{\label{fig:Model-of-SUT}Model of SUT Web Page and related concepts}
\end{figure*}

The model of the Web Page and related concepts are depicted in Figure
\ref{fig:Model-of-SUT}. The blocks with a white background depict
parts of the model that are automatically reengineered by the Tapir
framework during the exploratory testing process. Of these parts,
the elements specifically related to the interaction of the tester
with the SUT are depicted by the blocks with a dotted background.
The blocks with a blue-gray background depict metadata entered by
the Test Lead during the exploratory testing process.

As explained in Section \ref{subsec:Principle-of-the}, the model
is continuously built during the exploratory testing process. The
team of testers $\mathit{T}$ contribute to this process, and $\mathit{W_{T}}$
denotes the SUT pages explored by the whole team while $\mathit{W_{t}}$
denotes SUT pages explored by the tester $t\in T$. By \textcolor{black}{analogy,
$\mathit{L_{T}}$ ($A\mathit{_{T}}$) denotes the SUT link (action)
elements explored by the whole team and $\mathit{L_{t}}$ ($\mathit{A_{t}}$)
denotes link (action) elements explored by the tester $t\in T$.}

\textcolor{black}{By principle, a link or action element can be exercised
additional times during the test exploration process because a page
can be repeatedly visited. To capture this fact, $visits(w)_{t}$,
$visits(l)_{t}$ and $visits(a)_{t}$ denotes the number of visits
of page $w$, link element $l$ and action element $a$ by tester
$t$, respectively. Additionally, $visits(w)_{T}$, $visits(l)_{T}$
and $visits(a)_{T}$ denote the number of visits of the page $w$,
link element $l$ and action element $a$ by all testers in the testing
team $T$, respectively.}

\textcolor{black}{For each input element $i\in I$, the Test Lead
can define a set of ECs $EC(i)$, which determine the input test data
that shall be entered by the exploratory testers during the tests.
When these ECs are not defined, $EC(i)$ is empty. For each $ec(i)\in EC(i)$,
if $range(i)$ is an interval, then $\mathit{ec(i)}$ is a sub-interval
of $range(i)$; and if $range(i)$ is a set of discrete values, then
$\mathit{ec(i)}\in range(i)$. }

\textcolor{black}{ECs can be dynamically defined during the exploratory
testing pro}cess, and certain classes can be removed from the model
while other classes can be added to the model, with $\bigcap_{ec(i)\in EC(i)}ec(i)=\emptyset$
for each $i\in I$.

In addition, $data(i)_{t}$ ($data(i)_{T}$) denotes a set of test
data values that are entered to input element $i$ by tester $t$
(by all testers in testing team $T$) during the testing process.

A set of test data combinations that are entered by tester $t$ for
the input elements $I_{a}\subseteq I_{w}$ connected to the action
element $a$ is denoted as $data(I_{a})_{t}=\{(d_{1},...\,,d{}_{n})\mid d_{1}\in data(i_{1})_{t},\,...\,,\,d_{n}\in data(i_{n})_{t},\:\:i_{1}...i_{n}\in I_{a}\}$.
A set of test data combinations that are entered by all testers in
testing team $T$ in the input elements $I_{a}\subseteq I_{w}$ connected
to the action element $a$ is denoted as $data(I_{a})_{T}=\{(d_{1},...\,,d{}_{n})\mid d_{1}\in data(i_{1})_{T},\,...\,,\,d_{n}\in data(i_{n})_{T},\:\:i_{1}...i_{n}\in I_{a}\}$.

The Test Lead can also set a priority for selected elements of the
SUT model. This priority is denoted as $prio(\mathtt{\mathbb{\mathfrak{\mathfrak{\mathscr{X}}}}})$,
$prio(\mathfrak{\mathscr{X}})\in\left\{ 1...5\right\} $, where five
is the highest priority. $\mathtt{\mathbb{\mathfrak{\mathfrak{\mathscr{X}}}}}$
denotes particular web page $w\in W$, link element $l\in L$, or
action element $a\in A$.

The presented model is inspired by a web application model proposed
by Deutsch \textit{et al.} \cite{deutsch2007specification}. In our
SUT model, significant changes have been made. We kept and modified
the definitions of SUT and its web page $w$. In the main tuple that
defines the SUT model, we removed the database and states, and then
we distinguish the system inputs as input $\mathit{I}$, action $A$
and link $L$ elements. In the definition of the SUT web page $w$,
we distinguish the action $A_{w}$ and link $L_{w}$ elements. Consequently,
we defined the transitions to the following page according to the
action transition rule sets $\Theta_{w}$ and $\varPhi_{w}$. The
home page $w_{0}$ is taken from the original model. We also applied
a concept of the error page $w_{e}$; however, we define it differently. 

The remaining elements of the model (master pages, team of testers,
visits of particular SUT pages and their elements, ECs, test data
combinations and element priorities) are completely different and
were defined by new elements, and they capture specific features of
the exploratory testing problem.

\section{\label{sec:Generation-of-navigational}Generation of Navigational
Test Cases from the Model}

As previously explained, the Tapir framework generates high-level
navigational test cases that are aimed at guiding a group of exploratory
testers through the SUT. The primary purpose of these test cases is
to guide the tester in the SUT. The test cases are dynamically created
from the SUT model during the exploratory testing process.

\subsection{\label{subsec:Structure-of-the-navigational-test-case}Structure
of the Navigational Test Case}

The navigational test cases are dynamically constructed from the SUT
model for an individual tester $t\in T$ during the exploratory testing
process. The navigational test case is constructed for the actual
page $w\in W$ visited in the SUT and helps the tester determine the
next step in the exploratory testing process. The structure of the
navigational test case is described as follows.
\begin{enumerate}
\item Actual page $w\in W$ visited in the SUT.
\item $L_{w}$ (list of all link elements that lead to other SUT pages that
are accessible from the actual page). In this list, the following
information is given:
\begin{enumerate}
\item $L_{w}\cap L_{t}$ (links elements that lead to other SUT pages that
are accessible from the actual page previously visited by the particular
tester $t$),
\item $visits(l)_{t}$ for each $l\in L_{w}\cap L_{t}$,
\item $L_{w}\cap L_{T}$ (links elements that lead to other SUT pages that
are accessible from the actual page previously visited by all testers
in team $T$),
\item $visits(l)_{T}$ for each $l\in L_{w}\cap L_{T}$, and
\item $prio(l)$ and $prio(w_{l})$for each $l\in L_{w}$. Link $l$ leads
from the actual page $w$ to the page $w_{l}$.
\end{enumerate}
\item $A_{w}$ (list of all action elements that lead to other SUT pages
that are accessible from the actual page). In this list, the following
information is given:
\begin{enumerate}
\item $A_{w}\cap A_{t}$ (action elements that lead to other SUT pages that
are accessible from the actual page previously visited by the particular
tester $t$),
\item $visits(a)_{t}$ for each $a\in A_{w}\cap A_{t}$,
\item $A_{w}\cap A_{T}$ (action elements that lead to other SUT pages that
are accessible from the actual page previously visited by all testers
in the team $T$),
\item $visits(a)_{T}$ for each $a\in A_{w}\cap A_{T}$, and
\item $prio(a)$ for each $a\in A_{w}$.
\end{enumerate}
\item Five elements are suggested for exploration in subsequent test steps
for each of the navigational strategies assigned to tester $t$ by
the Test Lead. These elements are sorted by their ranking, which is
calculated by the respective navigational strategies. This ranking
gives the tester the flexibility to choose an alternative element
if he considers the suggested option to be unsuitable. Because a particular
tester can have additional navigational strategies available, these
suggestions are displayed for each of the assigned navigational strategies
in a separate list, and the tester can choose the optimal strategy
according to his personal testing strategy within the navigational
strategies set by the Test Lead. The elements that are suggested for
exploration are described as follows:
\begin{enumerate}
\item The link $l_{next}\in L_{w}$ and page $next(w)$ suggested for exploration
in the next test step, or
\item The action element $a_{next}\in A_{w}$ suggested for exploration
in the next test step.
\end{enumerate}
\item For each $a\in A_{w}$, if $I_{a}\neq\emptyset$: 
\begin{enumerate}
\item $data(I_{a})_{t}$,
\item $data(I_{a})_{T}$,
\item for each $i\in I_{a}$:
\begin{enumerate}
\item $ec(i)\in EC(i)$ is suggested based on the test data strategy (refer
to section \ref{subsec:Test-data-strategies}) set by the Test Lead,
and based on this suggestion, tester $t$ can select a particular
data value from $ec(i)$ to enter it into the input element $i$ for
the actual test;
\item $data(i)_{t}$ (all test data previously entered by the particular
tester $i$ to the input element $i$); and
\item $data(i)_{T}$ (all test data previously entered by all testers in
the testing team $T$ to the input element $i$).
\end{enumerate}
\end{enumerate}
\item Previous test data combinations entered in $I_{a}$, which resulted
in the error page $w_{e}$ (e.g., a J2EE exception page ) or a standardized
error message (typically, a PHP parsing error message or application
specific error message formatted in a unified standard manner) that
can be recognized by the Tapir framework.
\item Notes for testers, which can be entered by the Test Lead onto page
$w$, all link elements from $L_{w}$ and all action elements $A_{w}$.
The Test Lead can enter these notes as simple textfields (the notes
are not defined in the model in Section \ref{subsec:The-SUT-Model}).
\end{enumerate}

\subsection{\label{subsec:Navigational-strategies}Navigational Strategies}

To create navigational test cases during the exploratory testing process,
several navigational strategies can be employed. These strategies
are specified in Table \ref{tab:Navigational-strategies}. A navigational
strategy determines a principal method that can be applied by a tester
to explore the SUT. Most of the navigational strategies can be adjusted
using a particular ranking function as specified in Table \ref{tab:Ranks-used-in-navigational-strategies}.
The navigational strategies address the guided exploration of new
SUT functions for all testers individually or as a collaborative work
by the testing team. The same process is enhanced by navigation driven
by priorities of the SUT pages, links and action elements or by regression
testing for a defined historical period. This latter strategy is also
applicable to retests of defect fixes after a new SUT release.

The navigational strategy determines the SUT user interface elements
that are suggested for the SUT page $w$ in the navigational test
case (refer \ref{subsec:Structure-of-the-navigational-test-case}).
The input of this process is the application context (tester $t$
and related metadata) and actual state of the SUT model that has been
specified in sub-section \ref{subsec:The-SUT-Model}. By the rules
specified in Table \ref{tab:Navigational-strategies} and the ranking
functions specified in Table \ref{tab:Ranks-used-in-navigational-strategies},
a list of $l\in L_{w}$ and $a\in A_{w}$ , which are sorted by these
rules and functions, is created.

\begin{table*}
\caption{\label{tab:Navigational-strategies}Navigational strategies}
\centering{}%
\begin{tabular}{|c|>{\centering}p{8cm}|>{\centering}p{4.2cm}|>{\centering}p{3cm}|}
\hline 
Navigational strategy & Rules for element suggestion for page $w$ and tester $t$. 

Element$\varepsilon$ can be link element $l\in L_{w}$ or action
element $a\in A_{w}$.  & Ranking functions (see Table \ref{tab:Ranks-used-in-navigational-strategies})
used & Use case\tabularnewline
\hline 
\hline 
RANK\_NEW & $\varepsilon$ satisfying the following conditions: 

(1) $visits(\varepsilon)_{t}=0$, AND 

(2) ($\varepsilon$ has the highest \textit{ElementTypeRank}($\varepsilon$)
OR a page $w_{n}\in next(w)$ to which $\varepsilon$ leads has the
highest \textit{PageComplexityRank}($w_{n}$)), AND

(3) $\varepsilon\in w\in W\setminus M$ are preferred to $\varepsilon\in w\in M$ & \textit{ElementTypeRank} \textit{PageComplexityRank} & Exploration of new SUT functions\tabularnewline
\hline 
RANK\_NEW\_TEAM & As RANK\_NEW, (1) modified to: 

$visits(\varepsilon)_{T}$ has the minimal value among all $\varepsilon\in L_{w}$
and $\varepsilon\in A_{w}$.  & \textit{ElementTypeRank}

\textit{PageComplexityRank} & Exploration of new SUT functions\tabularnewline
\hline 
RT\_TIME & $\varepsilon$ satisfying the following conditions: 

(1) $visits(\varepsilon)_{t}>0$, AND 

(2) time elapsed from the last exploration of $\varepsilon$ by tester
$t$ \textgreater{} $LastTime$ constant, AND

(3) $\varepsilon\in w\in W\setminus M$ are preferred to $\varepsilon\in w\in M$ & - & (1) Retesting of defect fixes, (2) Regression testing\tabularnewline
\hline 
PRIO\_NEW & $\varepsilon$ satisfying the following conditions: 

(1) $visits(\varepsilon)_{t}=0$, AND 

(2) $prio(\varepsilon)$ has the maximal value among all $\varepsilon\in L_{w}$
and $\varepsilon\in A_{w}$, AND

(3) if $\varepsilon$ is a link element $l\in L_{w}$, page $w_{n}\in next(w)$
has the highest \textit{PriorityAndComplexityRank}($w_{n}$), AND

(4) $\varepsilon\in w\in W\setminus M$ are preferred to $\varepsilon\in w\in M$ & \textit{PriorityAndComplexityRank} & Exploration of new SUT functions by priorities set by the Test Lead\tabularnewline
\hline 
PRIO\_NEW\_TEAM & As RANK\_NEW, (1) modified to: 

$visits(\varepsilon)_{T}$ has the minimal value among all $\varepsilon\in L_{w}$
and $\varepsilon\in A_{w}$.  & \textit{PriorityAndComplexityRank} & Exploration of new SUT functions by priorities set by the Test Lead\tabularnewline
\hline 
\end{tabular}
\end{table*}

The ranking function \textit{ElementTypeRank} is used for both link
elements $l\in L_{w}$ and action elements $a\in A_{w}$. The \textit{PageComplexityRank}
is only used for link elements $l\in L_{w}$. In the case of action
elements, we are not able to determine the exact SUT page after the
process triggered by the action \textcolor{red}{element $a$} because
test data that have been entered can have a role in determining the
page that will be displayed in the next step (refer to the SUT model
in sub-section \ref{subsec:The-SUT-Model}).

\begin{table*}
\caption{\label{tab:Ranks-used-in-navigational-strategies}Ranks used in navigational
strategies}
\centering{}%
\begin{tabular}{|c|>{\centering}p{9cm}|}
\hline 
Rank & Definition\tabularnewline
\hline 
\hline 
\textit{ElementTypeRank}($\varepsilon$) & IF $\varepsilon$ is link THEN \textit{ElementTypeRank}($\varepsilon$)
= 1 

IF $\varepsilon$ is action element THEN \textit{ElementTypeRank}($\varepsilon$)
= 2\tabularnewline
\hline 
\textit{PageComplexityRank}($w_{n}$) & \textit{PageComplexityRank}($w_{n}$) = ((($\mid I_{w}\mid\cdotp inputElementsWeight$
+$\mid A_{w}\mid)\cdotp actionElementsWeight$ +$\mid L_{w}\mid$)
$\cdotp linkElementsWeight$

$I_{w}\in w_{n}$, $A_{w}\in w_{n}$, $L_{w}\in w_{n}$\tabularnewline
\hline 
\textit{PriorityAndComplexityRank}($w_{n}$) & \textit{PriorityAndComplexityRank}($w_{n}$) = (((($prio(w_{n})\cdotp pagePriorityWeight$
+$\mid I_{w}\mid$)$\cdotp inputElementsWeight$ +$\mid A_{w}\mid)\cdotp actionElementsWeight$
+$\mid L_{w}\mid$) $\cdotp linkElementsWeight$

$I_{w}\in w_{n}$, $A_{w}\in w_{n}$, $L_{w}\in w_{n}$\tabularnewline
\hline 
\end{tabular}
\end{table*}

In the \textit{ElementTypeRank}, action elements are preferred to
link elements because action elements can be expected to contain additional
business logic and data operations of the SUT to be explored in the
tests (typically, submitting the data by forms on SUT pages).

In the \textit{PageComplexityRank} ranking function, the constants
$actionElementsWeight$, $inputElementsWeight$, and $linkElementsWeight$
determine how strongly the individual page action elements, input
elements and link elements are preferred for determining the page
$w_{n}\in next(w)$. These elements are suggested for exploration
in the tester\textquoteright s next step via the use of a link element
that leads to $w_{n}$. An increase of a particular constant will
cause the pages with a higher number of particular elements to be
preferred. These constants can be dynamically set by the Test Lead
during the exploration testing process.

The constants $actionElementsWeight$ and $linkElementsWeight$ can
be set in the range of 1 to 512. The constant $inputElementsWeight$
can be set in the range of 0 to 512. The default value of these constants
is 256. Without any change, the pages with a higher number of forms,
a higher number of input fields, a higher number of action elements,
and a higher number of link elements are considered more complex for
the testing purposes, and they are suggested for initial exploration.

In the \textit{PriorityAndComplexityRank}, the priorities of the SUT
pages set by the Test Lead have the strongest role in the determination
of the suggested next page for exploration. The constant $pagePriorityWeight$
determines the extent of the role of this prioritization. Then, the
decision is influenced by the number of input fields, the number of
action elements, and the number of link elements. The constants $actionElementsWeight$,
$inputElementsWeight,$ and $linkElementsWeight$ have the same meaning
and function as in the \textit{PageComplexityRank}. The constant $pagePriorityWeight$
can be set in the range of 0 to 512 and its default value is 256.
In Tapir HQ component, the Test Lead is provided with a guideline
for setting the constants $actionElementsWeight$, $inputElementsWeight$,
$linkElementsWeight$ and $pagePriorityWeight$ and the influence
of a particular setting.

\subsection{\label{subsec:Test-data-strategies}Test Data Strategies}

During the construction of the navigational test cases, test data
are suggested for the input elements $I_{a}\subseteq I_{w}$ connected
to the action elements $a\in A_{w}$ of the particular page$w\in W$.
For this suggestion, (1) test data previously entered by the testers
($data(i)_{t}$ and $data(i)_{T}$ for each $i\in I_{a}$) and (2)
ECs defined by the Test Lead ( $EC(i)$ for each $i\in I_{a}$) are
employed.

For this process, the test data strategies described in Table \ref{tab:Test-data-strategies}
are available. These test data strategies are specifically designed
for different cases in the testing process, such as retesting defect
fixes, testing regressions or exploring new test data combinations.

\begin{table*}
\caption{\label{tab:Test-data-strategies}Test data strategies}
\centering{}%
\begin{tabular}{|c|>{\centering}p{8.5cm}|>{\centering}p{4cm}|}
\hline 
Test data strategy & Description & Use case\tabularnewline
\hline 
\hline 
DATA\_REPEAT\_LAST & For each $i\in I_{a}$, suggest the value of $data(i)_{t}$ used in
the last test made by tester $t$ on page $w$.

If $data(i)_{t}=\emptyset$, no suggestion is made.  & (1) Retesting of defect fixes, (2) Regression testing\tabularnewline
\hline 
DATA\_REPEAT\_RANDOM & Suggest a randomly selected test data combination from $data(I_{a})_{t}$.

If $data(I_{a})_{t}=\emptyset$, no suggestion is made.  & Regression testing\tabularnewline
\hline 
DATA\_REPEAT\_RANDOM\_TEAM & Suggest a randomly selected test data combination from $data(I_{a})_{T}$.

If $data(I_{a})_{T}=\emptyset$, no suggestion is made.  & Regression testing\tabularnewline
\hline 
DATA\_NEW\_RANDOM & For each $i\in I_{a}$: 

if $EC(i)\neq\emptyset$, suggest a $ec(i)\in EC(i)$, such that $d\notin ec(i)$
for any $d\in data(i)_{t}$

if $EC(i)=\emptyset$, suggest a value $d\in range(i)$, such that
$d\notin data(i)_{t}$ & Exploration of new test data combinations\tabularnewline
\hline 
DATA\_NEW\_RANDOM\_TEAM & For each $i\in I_{a}$: 

if $EC(i)\neq\emptyset$, suggest a $ec(i)\in EC(i)$, such that $d\notin ec(i)$
for any $d\in data(i)_{T}$

if $EC(i)=\emptyset$, suggest a value $d\in range(i)$, such that
$d\notin data(i)_{T}$ & Exploration of new test data combinations\tabularnewline
\hline 
DATA\_NEW\_GENERATED & The Tapir engine suggests combination, which was not used previously
by individual tester $t$. Combination of test data is taken from
a pipeline of test data combinations created by a Combination Interaction
Testing (CIT) module, connected to the framework by the defined interface
(details follow in section \ref{subsec:The-Tapir-architecture}). & Exploration of new test data combinations\tabularnewline
\hline 
DATA\_NEW\_GENERATED\_TEAM & As DATA\_NEW\_GENERATED\_TEAM, modified to: combination, which was
not used previously by any tester of the testing team $T$ & Exploration of new test data combinations\tabularnewline
\hline 
\end{tabular}
\end{table*}

Similar to the case of the navigational strategies, the test data
strategies for independent exploration of the SUT by individual testers
or team collaboration are available, and they are marked by the postfix
\textquotedblleft \_TEAM\textquotedblright{} in the name of the test
data strategy.

The test data strategy DATA\_NEW\_RANDOM\_TEAM aims to minimize the
particular duplicated test data variants entered by multiple testers
either by chance or by an improper work organization during the exploration
of new test data variants. Another case is intended for testing defect
fixes or regression testing, where DATA\_REPEAT\_LAST, DATA\_REPEAT\_RANDOM
and DATA\_REPEAT\_RANDOM\_TEAM strategies are available to reduce
a tester\textquoteright s overhead by remembering the last entered
test data. The team strategy DATA\_REPEAT\_RANDOM\_TEAM can improve
the efficiency of the process by minimizing particular duplicated
test data variants entered by multiple testers during regression testing.

The ECs entered by the Test Lead during his pioneering exploration
of the SUT prevent the input of test data that belong to one EC, which
exercises the same SUT behavior according to the SUT specification.

The possibility of connecting the Tapir framework to a CIT module
(DATA\_NEW\_GENERATED and DATA\_NEW\_GENERATED\_TEAM strategies) makes
the process more controlled and systematic. Only the efficient set
of test data combinations are used by the testers to exercise the
SUT functions.

\section{\label{sec:Experimental-Setup}Setup of the Case Studies}

To evaluate the proposed framework, compare its efficiency with the
manually performed exploratory testing process, and assess the proposed
navigational strategies, we conducted three case studies. These case
studies primarily focus on the process efficiency of the exploration
of new SUT functions, and their objective is to answer the following
research questions.

\textbf{\textit{RQ1}}: In which aspects the efficiency of the exploratory
testing process is increased by the Tapir framework when compared
with its manual performance? What is this difference?

\textbf{\textit{RQ2}}: Are there any aspects for which the Tapir framework
decreases the efficiency of the exploratory testing process compared
with its manual execution?

\textbf{\textit{RQ3}}: Which of the proposed navigational strategies
and ranking functions designed for exploration of new parts of the
SUT are the most efficient strategies and functions?

\textbf{\textit{RQ4}}: How the previous tester's experience influences
defects found in the exploratory testing process supported by the
Tapir framework when compared with its manual performance? 

Details of the case studies are presented in the following subsections.

\subsection{\label{subsec:The-Tapir-architecture}Tapir Architecture and Implementation}

As mentioned in Section \ref{subsec:Principle-of-the}, the Tapir
framework consists of three principal parts: Tapir Browser Extension,
Tapir HQ component, and Tapir Analytics module. A tester interacts
with the SUT in a browser window with an installed extension. In the
second window, the tester interacts with a Tapir HQ front-end application,
which serves as a test management tool. Here, suggestions for the
navigational test cases are presented to the tester. The Analytics
module can be accessed by the testers, Test Leads, and administrator
as a separate application and enables the visualization of the state
of the SUT model.

In this section, we provide additional implementation details of the
functionality of these framework modules.

\textbf{The Tapir Browser Extension} The Tapir Browser Extension is
implemented as a web browser plugin. The extension analyzes the current
page, intercepts the internal browser events (e.g., page was loaded
or redirected, user navigated back, or authentication is required),
and it registers event handlers for all links and buttons on the page.
All events relevant to the Tapir framework functionality are intercepted
and tracked. The browser extension has a functionality to highlight
SUT page elements in a mode selected by the Test Lead (elements already
analyzed by the Tapir framework or elements suggested for exploration
in the tester\textquoteright s next step). Currently, the browser
extension is developed for the Chrome browser to cover the highest
market share. The extension is implemented in JavaScript. Currently,
we are working on a Firefox version of the browser extension, which
can be simplified by browser extension portability\textcolor{black}{{}
}\footnote{\textcolor{black}{https://developer.mozilla.org/en-US/Add-ons/WebExtensions/
Porting\_a\_Google\_Chrome\_extension}}\textcolor{black}{. }

\textbf{The Tapir HQ} represents the core of the framework functionality.
This module receives the events from the browser extensions, constructs
the model, constructs the navigational test cases, and presents them
to the tester. Tapir HQ is a client application that is implemented
as a JavaScript single page application using the ReactJS framework.
The server back-end is implemented in .Net C\#. When a tester starts
testing the SUT, a socket is opened between the Tapir HQ back-end
service and Tapir HQ front-end application in the browser to synchronize
the data in real time. For this communication, the SocketIO library
is employed. Tapir HQ also contains an open interface to a CIT module
to import preferred test data combinations (when test data strategies
DATA\_NEW\_GENERATED and DATA\_NEW\_GENERATED\_TEAM are employed;
refer to Table \ref{tab:Test-data-strategies}). The interface is
based on uploading CSV files of a defined structure or a defined JSON
format. The test data combinations are determined for the input elements
$I_{a}\subseteq I_{w}$ connected to the action elements $a\in A_{w}$
of the particular page$w\in W$. For the action element $a\in A_{w}$,
the inputs to the CIT module are $I_{a}$ and $EC(i)$ for each $i\in I_{a}$.
The output from the CIT module is a set of test data combinations.
The data combination $(d_{i_{1}},...,d_{i_{n}})$ is a set of values
to be entered in $i_{1}...i_{n}\in I_{a}$. The test data combinations
are stored in a pipeline and suggested to the testers to be entered
in $I_{a}$ during the tests. In cases in which $EC(i)$ is not defined
by the Test Lead, a random value from $range(i)$ is employed. In
this case, the test data combinations are marked by a special flag.
For the generation of the test data combinations, PSTG (Particle Swarm
Test Generator) algorithm \cite{ahmed2012application} is applied.
The process of generating test data combinations can be run by the
Test Lead or system administrator for $I_{a}$ when $EC(i)$ for $i\in I_{s}\subseteq I_{a}$.
are defined. If previous test data combinations are available, they
are overridden by the new set. 

To store the SUT model, the MongoDB NoSQL database is employed. The
document-oriented NoSQL database was selected because of its efficiency
for JSON document processing. Documents can be directly stored in
this database. The database with the SUT model is shared by both the
Tapir HQ module and Tapir Analytics module. The Tapir HQ back-end
service exposes the API to access the database by the individual modules.
In the current version of the framework, user authorization is implemented
via Google authentication, which is supported by the Chrome browser.

\textbf{The Tapir Analytics} module enables users to visualize the
current state of the SUT model. The visualization is in the form of
a textual representation or a directed graph and a particular state
of SUT exploration. This part is implemented as a module of Tapir
HQ that shares the SUT model with the Tapir HQ application. The framework
administrator grants access rights to this module. This part is implemented
in .Net C\#. Visualization of the SUT model is implemented in the
ReactJS framework.

\begin{figure*}
\centering{}\includegraphics[scale=0.6]{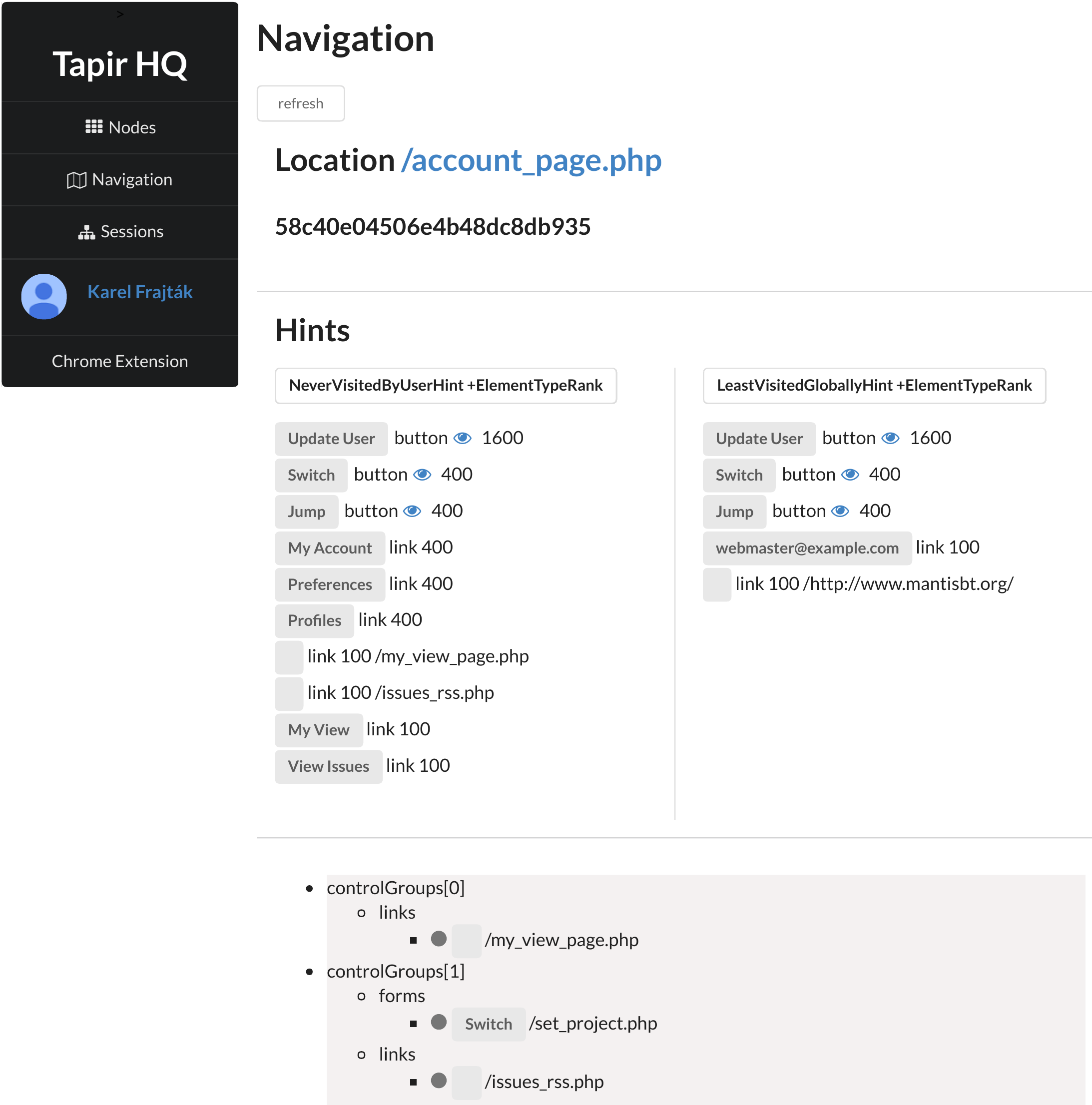}\caption{\label{fig:A-sample-of-testers-screen}A sample of testers' navigational
support (simplified)}
\end{figure*}

\textcolor{black}{Figure \ref{fig:A-sample-of-testers-screen} depicts
a tester\textquoteright s navigation support in the SUT via Tapir
HQ. The system displays suggested actions to be explored (sorted by
ranking functions) by several navigational strategies available to
the user. Regarding the format of this article, the view is simplified:
test data suggestions and other details are not visible in this sample.}

\begin{figure*}
\begin{centering}
\includegraphics[scale=0.5]{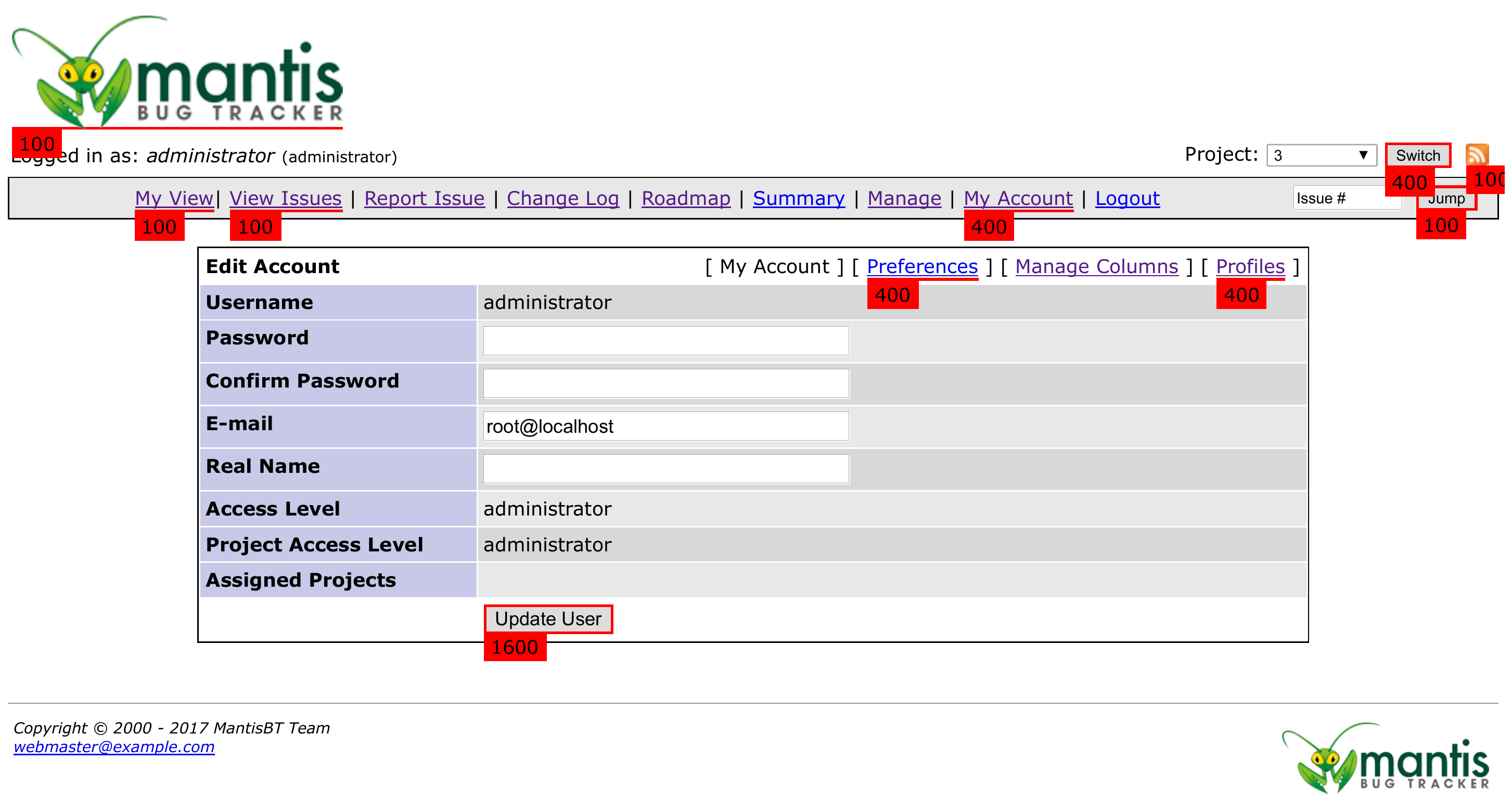}
\par\end{centering}
\caption{\label{fig:SUT-with-the-browser-extension-highlights}S\textcolor{black}{UT
with the Tapir Browser Extension highlights}}

\end{figure*}

\textcolor{black}{In Figure \ref{fig:SUT-with-the-browser-extension-highlights},
a corresponding screenshot from the SUT is presented. The Tapir Browser
Extension highlights the elements to be explored in the next step
and displays the value of the ranking function for these elements.}

\begin{figure*}
\begin{centering}
\includegraphics[scale=0.6]{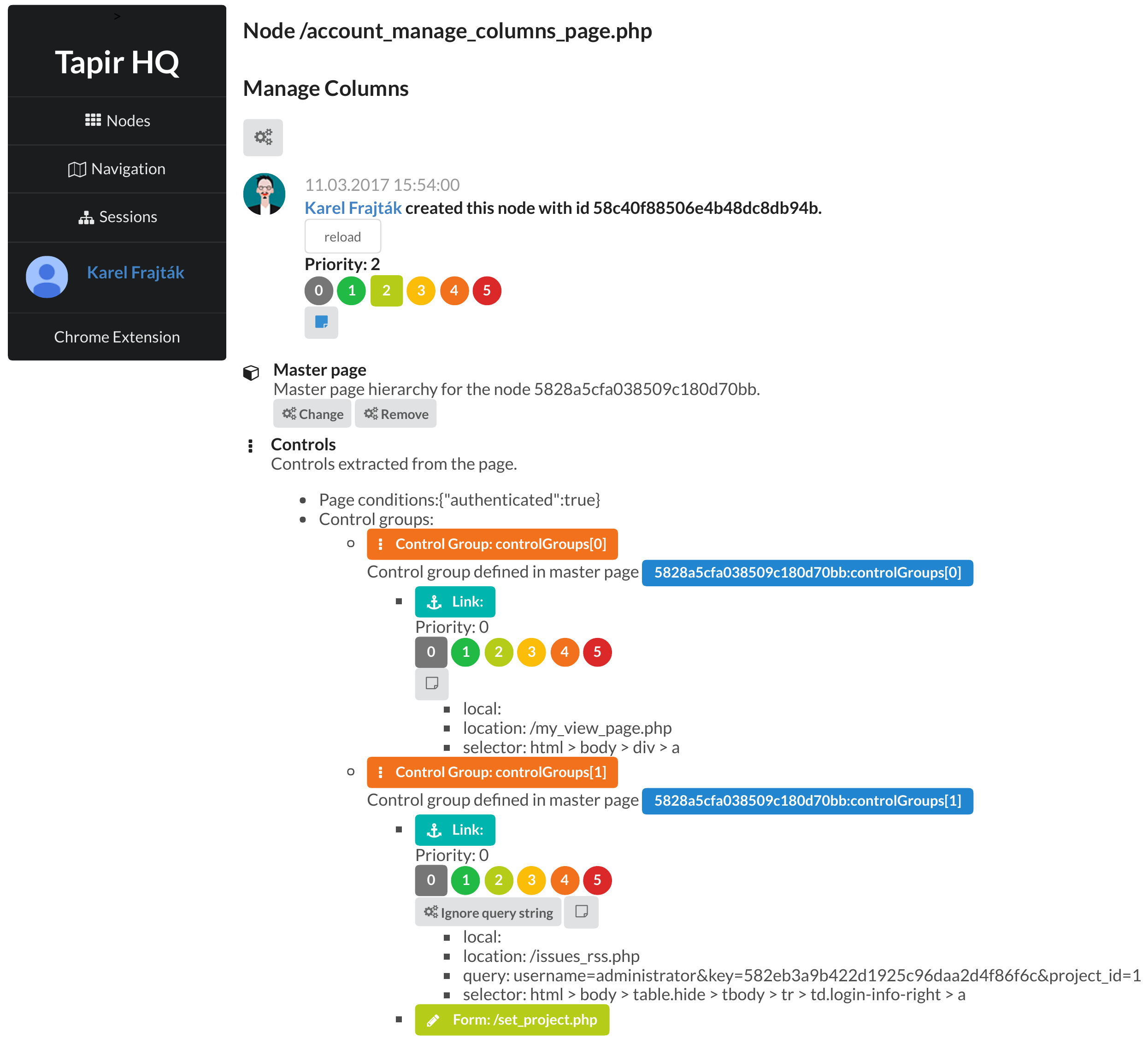}
\par\end{centering}
\caption{\label{fig:A-sample-of-test-leads-screen}A sample of Test Lead's
administration of particular SUT page}
\end{figure*}

Figure \ref{fig:A-sample-of-test-leads-screen} depicts a sample of
the Test Lead\textquoteright s administration of particular SUT page.
Prioritization of the SUT pages and elements can be performed in this
function. 

\begin{figure*}
\begin{centering}
\includegraphics[width=18cm]{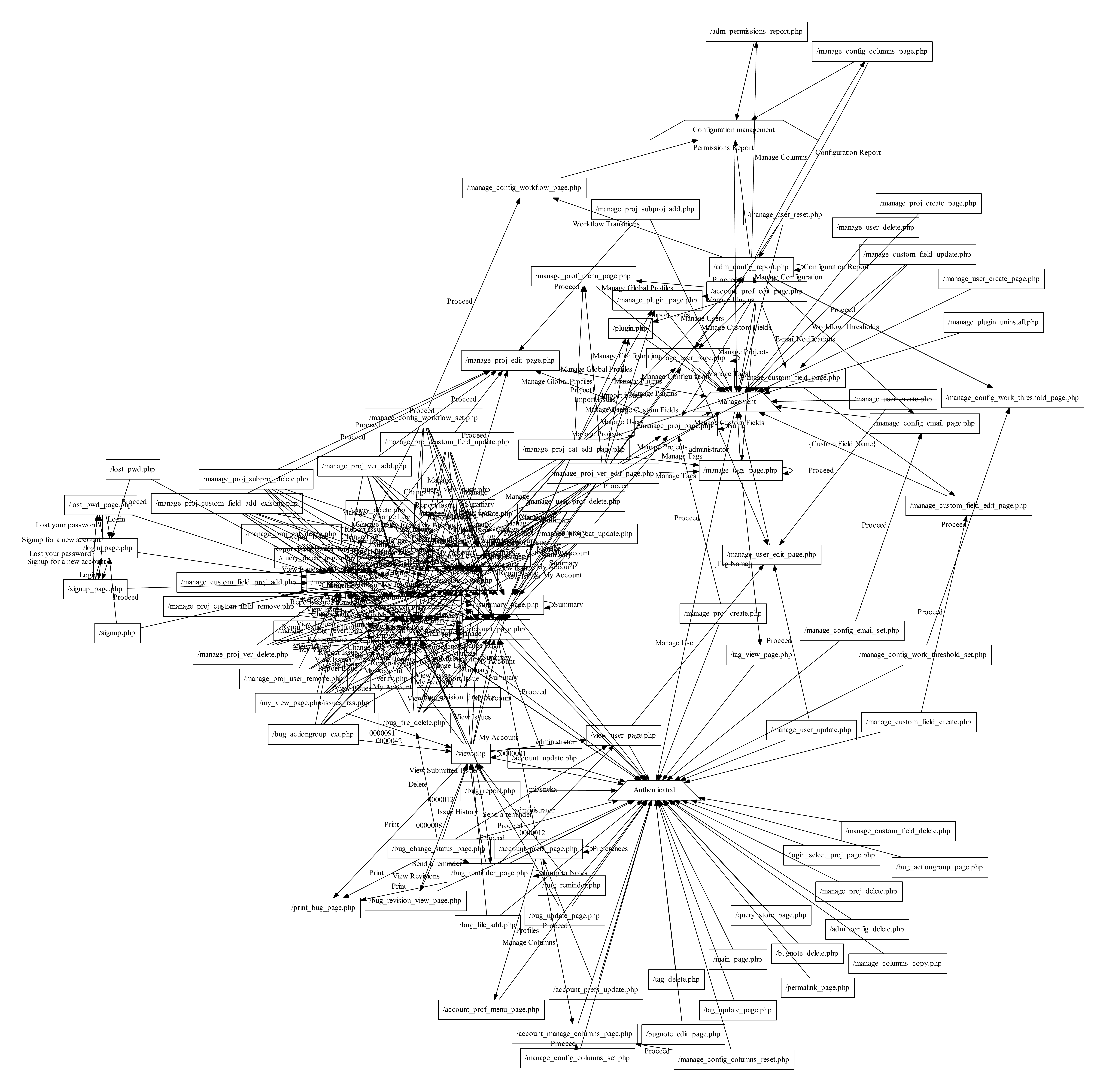}
\par\end{centering}
\caption{\label{fig:A-sample-from-analytics-module}A sample from Analytics
module}
\end{figure*}

Figure \ref{fig:A-sample-from-analytics-module} depicts a sample
from the Analytics module, which consists of the visualization of
SUT pages and possible transitions between the pages. Because the
graph is usually extensive, it can be arranged by several layouts
to obtain additional user comfort. In the sample, compact visualization
is depicted.

\subsection{\label{subsec:System-Under-Test}Systems Under Test with Injected
Defects}

In Case Studies 1 and 2, we employed an open-source MantisBT\footnote{https://www.mantisbt.org/}
issue tracker as a SUT for the experiments. The MantisBT is written
in PHP and uses a relational database. We modified the source code
of the SUT by inserting 19 artificial defects. We accompanied the
defective code lines by a logging mechanism for reporting each activation
of the defective line code. The details of the injected defects are
illustrated in Table \ref{tab:Defects-injected-to}.

\begin{table*}
\caption{\label{tab:Defects-injected-to}Defects injected to the system under
test}
\centering{}%
\begin{tabular}{|c|c|>{\centering}p{8cm}|}
\hline 
Injected defect ID & Type & SUT function\tabularnewline
\hline 
\hline 
synt\_1 & Syntax error & Plugin installation function broken\tabularnewline
\hline 
synt\_2 & Syntax error & Plugin uninstallation function broken\tabularnewline
\hline 
synt\_3 & Syntax error & Import issues from XML function broken\tabularnewline
\hline 
synt\_4 & Syntax error & Adding empty set of users to a project causes system defect of the
SUT\tabularnewline
\hline 
synt\_5 & Syntax error & Setting configuration option with empty value causes system defect
of the SUT\tabularnewline
\hline 
synt\_6 & Syntax error & Config option of float type cannot be created\tabularnewline
\hline 
synt\_7 & Syntax error & Config option with complex type cannot be created\tabularnewline
\hline 
mc\_1 & Missing code & Export to CSV is not implemented\tabularnewline
\hline 
mc\_2 & Missing code & The action ``set sticky'' in search issues screen is not implemented\tabularnewline
\hline 
mc\_3 & Missing code & Printing of the issue details is not implemented\tabularnewline
\hline 
mc\_4 & Missing code & User cannot be deleted\tabularnewline
\hline 
mc\_5 & Missing code & Bug note cannot be deleted\tabularnewline
\hline 
cc\_1 & Change in condition & Issue configuration option value cannot be set in database\tabularnewline
\hline 
cc\_2 & Change in condition & Issue configuration option value in not loaded properly from database\tabularnewline
\hline 
cc\_3 & Change in condition & Tag with the name \char`\"{}Tapir\char`\"{} (predefined in the SUT)
cannot be deleted\tabularnewline
\hline 
var\_1 & Wrong set of variable & Language in user preferences is always \char`\"{}English\char`\"{}
and cannot be changed\tabularnewline
\hline 
var\_2 & Wrong set of variable & User defined columns in issue list cannot be copied between projects\tabularnewline
\hline 
var\_3 & Wrong set of variable & When adding new bug note, its status cannot be \char`\"{}private\char`\"{}\tabularnewline
\hline 
var\_4 & Wrong set of variable & Bug note view status cannot be changed\tabularnewline
\hline 
\end{tabular}
\end{table*}

In Case Study 3, we used the healthcare information system Pluto that
was developed in a recent software project\footnote{http://www.lekis.cz/Stranky/Reseni.aspx}.
The Pluto system is employed by hospital departments for a complete
supply workflow of pharmaceutical products and medical equipment.
The system is one part of a complex hospital information system. The
front-end of the Pluto system is created using HTML, CSS and the JavaScript
framework Knockout. The back-end of the system is developed in C\#
and runs on the .NET platform. In this case, we obtained a history
of real defects that were detected and fixed in the software from
December 2015 to September 2017. As a SUT for the experiments, we
used a code baseline of the system from August 2015. We analyzed 118
defects that were reported from December 2015 to September 2017. During
the bug fixing and maintenance phase of the SUT, 72 of the defects
were found to be caused by the regression as a result of implemented
change requests and other bug fixes. During the initial analysis,
we excluded these defects; thus, the final number of defects in the
experiment was \textcolor{red}{48}.

We obtained information about the presence of the defects in the SUT
code. We accompanied the SUT with a .NET Aspect-based logging mechanism
to log the flows calls of the SUT methods during the tests. Using
this mechanism, we were able to determine and analyze the defects
in the code that were activated by the exercised tests. We automated
this analysis by a set of scripts that compare the data of the SUT
model recorded by the Tapir framework with these defect activation
logs.

\subsection{Setup of Case Studies}

In Case Study 1, we compared the exploratory testing process that
was manually performed by individual testers with the exploratory
testing process supported by the Tapir framework presented in this
paper. The aim of this case study is to answer research questions
\textit{RQ1} and \textit{RQ2}. In this case study, we employed the
following method.

A group of 54 testers performed exploratory testing in the SUT. The
MantisBT issue tracker and inserted artificial defects are employed
(refer to Table \ref{tab:Defects-injected-to}). Each of the testers
were allowed to individually act, and they were instructed to perform
an exploratory smoke test and explore the maximal extent of the SUT.
Exit criteria were left for an individual tester\textquoteright s
consideration.

To evaluate the results of this case study, we applied data that were
available in the SUT model created by the Tapir framework during the
exploratory testing process (for details, refer to Section \ref{subsec:Results-of-the-case-study-1}).
A subjective report by individual testers was not used in the evaluation.
The testers were divided into two groups.
\begin{enumerate}
\item \textcolor{black}{A group of 23 testers manually performed the exploratory
testing process. The activity of these testers was recorded by the
Tapir framework tracking extension and Tapir HQ Back-End service.
The Tapir HQ Front-End application was not available to this group.
Thus, navigational support was not provided to its members.}
\item A group of 31 testers disjunctive to the previous group performed
the exploratory testing process with support provided by the Tapir
framework. This group employed the RANK\_NEW navigational strategy.
Within this strategy, half of this group is randomly selected to use
\textit{PageComplexityRank} and the other half of the group is selected
to use \textit{ElementTypeRank}. Although DATA\_NEW\_RANDOM was employed
as the test data strategy, the Team Lead did not define any ECs. The
testers in this group were explicitly instructed to not use the test
data suggestions made by the framework, and their task was to determine
which test data to actively enter. The purpose of this task was to
equalize the conditions of both groups (the group that manually performs
the exploratory testing has no support regarding the test data). No
priorities were set for the SUT pages and its elements. The Test Lead
did not change any setup during the experiments. The values of the
$actionElementsWeight$, $inputElementsWeight,$ and $linkElementsWeight$
constants were set to the default value 256.
\end{enumerate}
The participants were differing in praxis in software testing from
0.5 to 4 years. The participants were randomly distributed in the
groups. In this case study, we have not employed the team variants
of the provided navigational strategies (RANK\_NEW\_TEAM). In an objective
experiment, equivalent team support must be provided for cases in
which exploratory testing is manually performed. We have attempted
to perform this initial experiment, and an equivalent simulation of
the Tapir framework functionality by a human team leader was difficult
to achieve. Thus, we evaluate the team versions of the navigational
strategies in Case Study 2. 

In Case Study 2, we focused on answering the research question \textit{RQ3}.With
an independent group of testers, we compared the proposed navigational
strategies that primarily focused on exploring new SUT functions.
In this study, a group of 48 testers performed exploratory testing
in the MantisBT issue tracker with inserted artificial defects (refer
to Table \ref{tab:Defects-injected-to}). All testers used the support
of the Tapir framework. The testers were instructed to explore the
maximal extent of the SUT. This group was split into four subgroups
as specified in Table \ref{tab:Participant-groups-performing-CS-2}.

\begin{table*}
\caption{\label{tab:Participant-groups-performing-CS-2} Participant groups
performing the Case Study 2}
\centering{}%
\begin{tabular}{|c|c|c|c|c|}
\hline 
Group ID & Number of participants & Navigational strategy & Ranking function & Test data strategy\tabularnewline
\hline 
\hline 
1 & 13 & RANK\_NEW\_TEAM & \textit{ElementTypeRank} & DATA\_NEW\_RANDOM\_TEAM\tabularnewline
\hline 
2 & 11 & RANK\_NEW & \textit{ElementTypeRank} & DATA\_NEW\_RANDOM\tabularnewline
\hline 
3 & 12 & RANK\_NEW\_TEAM & \textit{PageComplexityRank} & DATA\_NEW\_RANDOM\_TEAM\tabularnewline
\hline 
4 & 12 & RANK\_NEW & \textit{PageComplexityRank} & DATA\_NEW\_RANDOM\tabularnewline
\hline 
\end{tabular}
\end{table*}

The participants were differing in praxis in software testing from
0.5 to 4 years. The participants were randomly distributed in the
groups. In this case study, ECs were not defined by the Team Lead.
In addition, the testers in this group were explicitly instructed
to ignore the test data suggestions made by the framework, and their
task was to determine which test data to actively enter. No priorities
were set for the SUT pages and its elements. The Test Lead did not
change any of the setup parameters during the experiment. The values
of the $actionElementsWeight$, $inputElementsWeight,$ and $linkElementsWeight$
constants were set to the default value 256.

Regarding the strategies that employ prioritization of the page elements
and pages (particularly, PRIO\_NEW and PRIO\_NEW\_TEAM), this concept
adds extra opportunities to improve the efficiency of the exploratory
testing process. A comparable alternative for the proposed navigational
strategies is not available at this developmental stage of the Tapir
framework. When equivalent prioritization is performed in the manual
exploratory testing process, we expect the same increase in testing
process efficiency. For these reasons, we have decided to exclude
the evaluation of the navigational strategies PRIO\_NEW and PRIO\_NEW\_TEAM
from the described case study.

Regarding the test data strategies, each of the individual strategies
is practically designed for different use cases (refer to Table \ref{tab:Test-data-strategies}).
A comparison can be performed between the strategies designed for
an individual tester\textquoteright s guidance and the strategies
designed for team exploratory testing, e.g., DATA\_NEW\_RANDOM versus
DATA\_NEW\_RANDOM\_TEAM. Because the presented case study primarily
focuses on the efficiency of process exploration for the new SUT functions,
a comparison of the strategies DATA\_NEW\_RANDOM and DATA\_NEW\_RANDOM\_TEAM
was included in Case Study 2.

In Case Study 3, we compared the exploratory testing process that
was manually performed by individual testers with the exploratory
testing process supported by the Tapir framework. In this case study,
another SUT with real software defects was employed (refer to Section\ref{subsec:System-Under-Test}).
This case study aims to answer research questions \textit{RQ1}, \textit{RQ2
and RQ4}. The organization of the experiment is described as follows.

A group of 20 testers performed exploratory testing in the Pluto system,
with each tester acting individually. The instructions to perform
an exploratory smoke test and explore the maximal extent of the SUT
were the same instructions provided in Case Study 1. Exit criteria
were left for an individual tester\textquoteright s consideration.

A group of ten testers performed the manual exploratory testing process,
and their activities were recorded by the Tapir framework. Navigational
support of the Tapir HQ was not provided to these testers. Another
group of ten testers employed the Tapir framework support. RANK\_NEW
was used as a navigational strategy. In this group, one randomly selected
half of the testers used \textit{PageComplexityRank}, whereas the
other half of the testers used \textit{ElementTypeRank}. Regarding
the testing data, we selected the same setup as in Case Study 1 to
ensure that the conditions of both groups were as equal as possible.
Although DATA\_NEW\_RANDOM was utilized, ECs were not defined by the
Test Lead, and the testers were instructed to define their individual
testing data (test data suggestions of the Tapir framework were not
employed). Element priorities were not applied, and the values of
the $actionElementsWeight$, $inputElementsWeight,$ and $linkElementsWeight$
constants were set to the default value 256. For objectivity reasons,
no team variant of a navigational strategy was utilized (this experiment
was the subject of Case Study 2). Experience of the testers varied
from 0.5 years to 3 years. We mixed both groups to have the average
experience of the testers in each of the groups 1.5.

Regarding the experimental groups, we ensured that all participants
had received the equivalent initial training regarding the following
software testing techniques: (1) principle of exploratory testing,
(2) identification of boundary values, (3) equivalence partitioning,
(4) testing data combinations to input in the SUT (condition, decision
and condition/decision coverage, pairwise testing and basics of constraint
interaction testing) and (5) techniques to explore a SUT workflow
(process cycle test). 

To evaluate the case studies, we used a set of metrics that are based
on the SUT model and activated defects in the SUT code. This set of
metrics is defined in Table \ref{tab:Metrics-used-to}.

\begin{table*}
\begin{centering}
\begin{tabular}{|c|c|c|}
\hline 
Metric definition & Explanation & Unit\tabularnewline
\hline 
\hline 
$\mid T\mid$  & Number of participants & -\tabularnewline
\hline 
\textit{pages} = $\sum_{w\in W}visits(w)_{T}$  & Total number of explored pages, pages can repeat & -\tabularnewline
\hline 
\textit{u\_pages} = $\mid W\mid$ & Total number of explored unique pages & -\tabularnewline
\hline 
\textit{r\_pages} = $\frac{\mid W\mid}{\sum_{w\in W}visits(w)_{T}}\cdot100\%$  & Ratio of explored unique pages  & \%\tabularnewline
\hline 
\textit{links} = $\sum_{l\in L}visits(l)_{T}$  & Total number of link elements explored, elements can repeat & -\tabularnewline
\hline 
\textit{u\_links} = $\mid L\mid$ & Total number of explored unique link elements & -\tabularnewline
\hline 
\textit{r\_links} = $\frac{\mid L\mid}{\sum_{l\in L}visits(l)_{T}}\cdot100\%$  & Ratio of explored unique link elements & \%\tabularnewline
\hline 
\textit{actions} = $\sum_{a\in A}visits(a)_{T}$   & Total number of action elements explored, elements can repeat & -\tabularnewline
\hline 
\textit{u\_actions} = $\mid A\mid$  & Total number of explored unique action elements & -\tabularnewline
\hline 
\textit{r\_actions} = $\frac{\mid A\mid}{\sum_{a\in A}visits(a)_{T}}\cdot100\%$  & Ratio of explored unique action elements & \%\tabularnewline
\hline 
\textit{time\_page = }$\frac{\tau}{\sum_{w\in W}visits(w)_{T}}$  & Average time spent on page & seconds\tabularnewline
\hline 
\textit{time\_u\_page = }$\frac{\tau}{\mid W\mid}$  & Average time spent on unique page & seconds\tabularnewline
\hline 
\textit{time\_link =} $\frac{\tau}{\sum_{l\in L}visits(l)_{T}}$  & Average time spent on link element & seconds\tabularnewline
\hline 
\textit{time\_u\_link =} $\frac{\tau}{\mid L\mid}$  & Average time spent on unique link element & seconds\tabularnewline
\hline 
\textit{time\_action = }$\frac{\tau}{\sum_{a\in A}visits(a)_{T}}$  & Average time spent on action element & seconds\tabularnewline
\hline 
\textit{time\_u\_action = }$\frac{\tau}{\mid A\mid}$  & Average time spent on unique action element & seconds\tabularnewline
\hline 
\textit{defects} & Average activated defects logged, activated defects can repeat & -\tabularnewline
\hline 
\textit{u\_defects} & Average unique activated defects logged & -\tabularnewline
\hline 
\textit{time\_defect} & Average time to activate one defect, activated defects can repeat & seconds\tabularnewline
\hline 
\textit{time\_u\_defect} & Average time to activate one unique defect & seconds\tabularnewline
\hline 
\end{tabular}
\par\end{centering}
\caption{\label{tab:Metrics-used-to}Metrics used to evaluate Case Studies
1-3}

\end{table*}

In the definitions, $\tau$ represents the total time spent by exploratory
testing activity, and it was averaged for all testers in the group
and given in seconds. The average time spent on a page is measured
using the Tapir framework logging mechanism. The average time to activate
a defect (\textit{time\_defect}) is calculated as the average total
time spent by the exploratory testing process divided by the number
of activated defects. When a defect is activated, it occurs during
the exploratory testing process. Thus, a tester can notice and report
this defect.

\section{\label{sec:Experiment-Results-and}Case Study Results}

In this section, we present and discuss the results of the performed
case studies.

\subsection{\label{subsec:Results-of-the-case-study-1}The Results of Case Study
1}

Table \ref{tab:Comparison-TAPIR-vs-MANUAL-data} summarizes the comparison
between the manual exploratory testing approach and the Tapir framework.
The results are based on the data that we were able to automatically
collect from the recorded SUT model. In this comparison, the average
of the results from the navigational strategy RANK\_NEW and ranking
functions \textit{ElementTypeRank} and \textit{PageComplexityRank}
are provided for the Tapir framework. DATA\_NEW\_RANDOM was utilized
as the test data strategy. $\mathit{DIFF=(AUT-MAN)/AUT}$ is presented
as a percentage, where $\mathit{AUT}$ represents the value measured
in the case of the Tapir framework, and $\mathit{MAN}$ denotes the
value measured in the case of the manual approach. In Table \ref{tab:Comparison-TAPIR-vs-MANUAL-data},
we use the metrics previously defined in Table \ref{tab:Metrics-used-to}.
In the statistics, we excluded excessive lengthy steps (tester spent
more than 15 minutes on a particular page) caused by leaving the session
open and not testing. In the case of the manual exploratory testing,
these excluded steps represented 1.19\% of the total recorded steps;
and in the case of Tapir framework support, this ratio was 0.72\%.

\begin{table*}
\caption{\label{tab:Comparison-TAPIR-vs-MANUAL-data} Comparison of manual
exploratory testing approach with Tapir framework: data from SUT model
for Case Study 1}
\centering{}%
\begin{tabular}{|c|c|c|c|}
\hline 
Metric & Manual approach & Tapir framework used & $\mathit{DIFF}$\tabularnewline
\hline 
\hline 
$\mid T\mid$ & 23 & 31 & -\tabularnewline
\hline 
\textit{pages} & 151.8 & 197.9 & 23.3\%\tabularnewline
\hline 
\textit{u\_pages} & 22.2 & 37.7 & 41.0\%\tabularnewline
\hline 
\textit{r\_pages} & 14.6\% & 19.0\% & 23.1\%\tabularnewline
\hline 
\textit{links} & 64.7 & 113.2 & 42.9\%\tabularnewline
\hline 
\textit{u\_links} & 21.4 & 44.0 & 51.3\%\tabularnewline
\hline 
\textit{r\_links} & 33.1\% & 38.9\% & 14.8\%\tabularnewline
\hline 
\textit{actions} & 24.5 & 59.0 & 58.5\%\tabularnewline
\hline 
\textit{u\_actions} & 9.6 & 28.3 & 66.2\%\tabularnewline
\hline 
\textit{r\_actions} & 39.1\% & 47.9\% & 18.5\%\tabularnewline
\hline 
\hline 
\textit{time\_page} & 21.5 & 20.1 & -6.6\%\tabularnewline
\hline 
\textit{time\_u\_page} & 146.7 & 105.7 & -38.7\%\tabularnewline
\hline 
\textit{time\_link} & 50.4 & 35.2 & -43.2\%\tabularnewline
\hline 
\textit{time\_u\_link} & 152.3  & 90.6 & -68.1\%\tabularnewline
\hline 
\textit{time\_action} & 133.1 & 67.5 & -97.3\%\tabularnewline
\hline 
\textit{time\_u\_action} & 340.7 & 140.8 & -141.9\%\tabularnewline
\hline 
\hline 
\textit{defects} & 11.1  & 16.6  & 32.8\%\tabularnewline
\hline 
\textit{u\_defects} & 4.6 & 6.6 & 31.1\%\tabularnewline
\hline 
\textit{time\_defect} & 292.8 & 240.5 & -21.7\%\tabularnewline
\hline 
\textit{time\_u\_defect} & 713.8 & 601.3 & -18.7\%\tabularnewline
\hline 
\end{tabular}
\end{table*}

To measure the activated defects, we accompanied the defective code
lines by a logging mechanism and reported each activation of the defective
line code.

Of the 19 inserted artificial defects, three defects, \textit{synt\_6},
\textit{synt\_7} and\textit{ mc\_2}, were not activated by any of
the testers in the group supported by the Tapir framework, which yields
a 15.8\% ratio. In the case of the manually performed exploratory
testing, the defect \textit{mc\_2} was activated by one tester from
the group. 

\begin{figure*}
\centering{}\includegraphics[width=18cm]{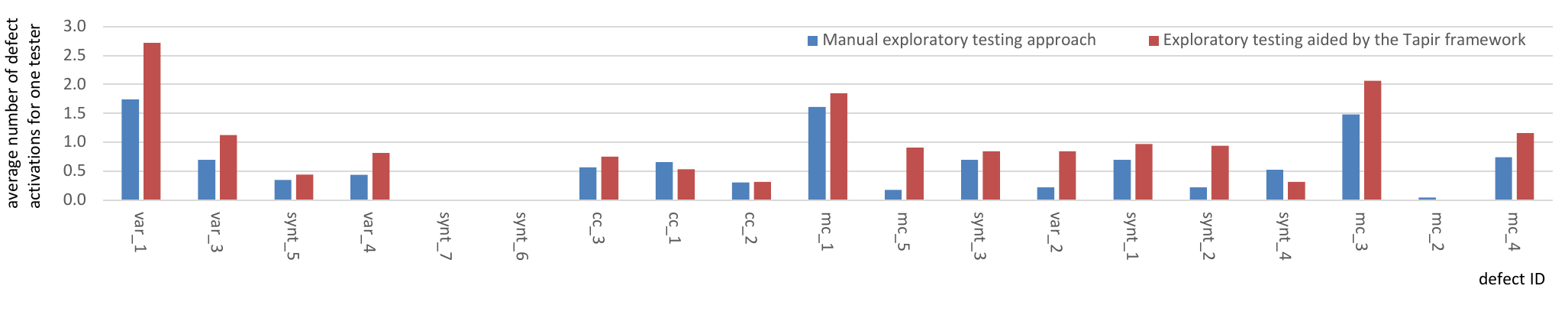}\caption{\label{fig:Potential-to-detect-defects}Potential of manual exploratory
testing and Tapir framework approach to detect injected defects in
the SUT in Case Study 1}
\end{figure*}

A comparison was performed between the manual exploratory testing
approach and the exploratory testing approach supported by the Tapir
framework to determine the efficiency of the potential to detect injected
artificial defects in the SUT, and it is depicted in Figure \ref{fig:Potential-to-detect-defects}.
For particular injected defects, an average value for the number of
times one tester activated the defect is presented. Value 1 indicates
that all the testers in the group have activated the defect once.
For example, value 0.5 indicates that 50\% of the testers in the group
have activated the defect once. The injected artificial defects are
introduced in Table \ref{tab:Defects-injected-to}.

\begin{figure*}
\begin{centering}
\includegraphics[width=18cm]{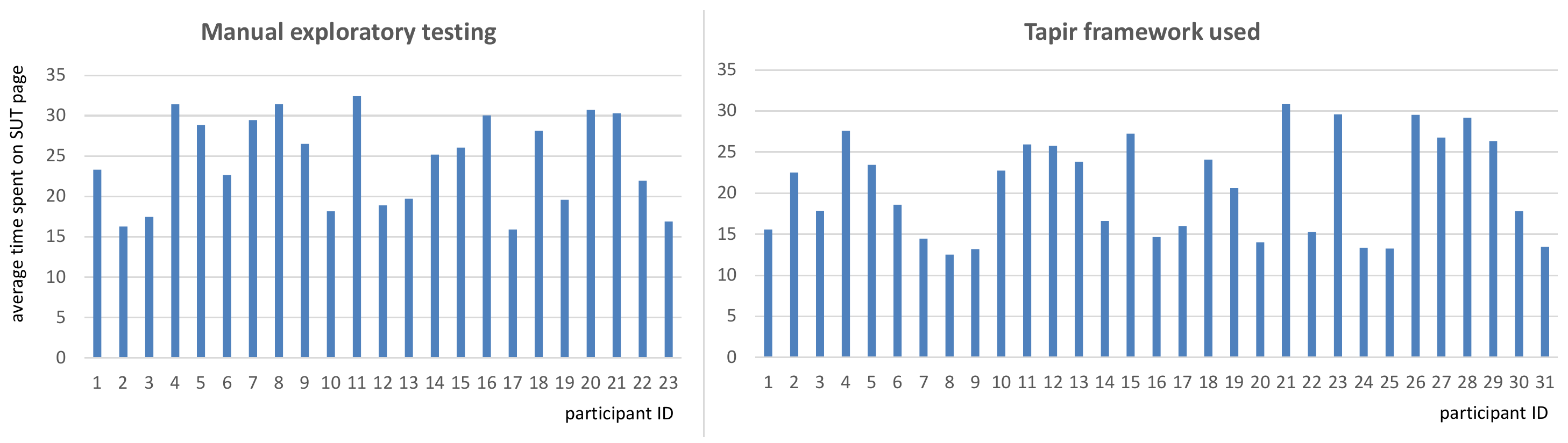}
\par\end{centering}
\caption{\label{fig:Average-times-spent}Average times spent on SUT pages by
testers using manual approach and Tapir framework}
\end{figure*}

Figure \ref{fig:Average-times-spent} provides details on the average
time spent on a SUT page by testers using the manual exploratory testing
approach and testers using the Tapir framework support. 

\begin{figure*}
\begin{centering}
\includegraphics[width=18cm]{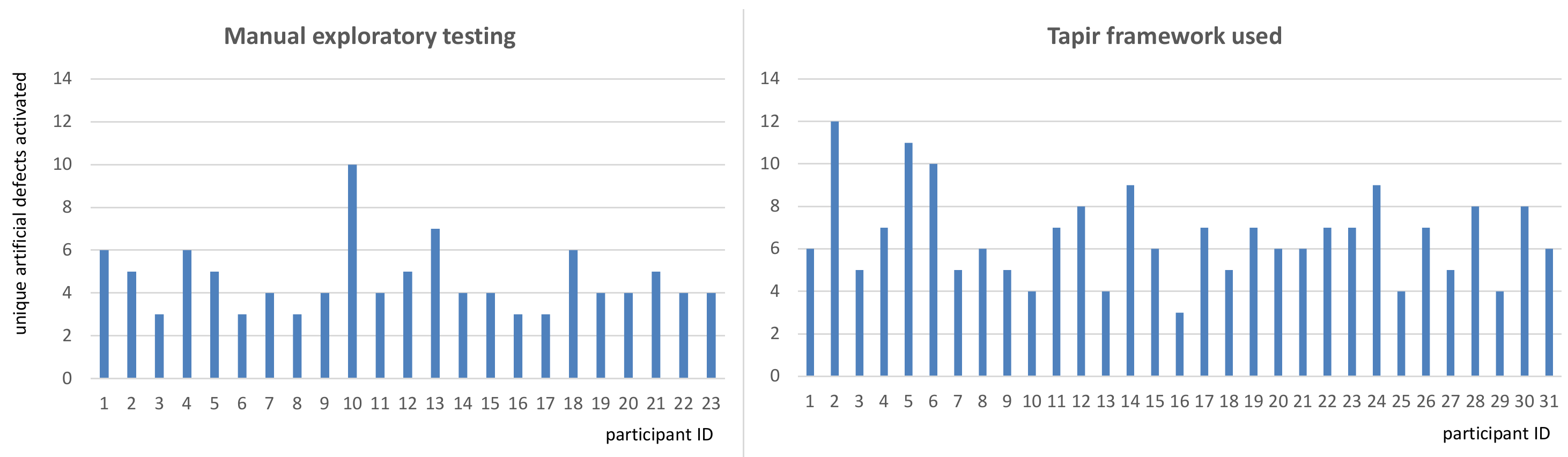}
\par\end{centering}
\caption{\label{fig:Unique-inserted-defects}Unique inserted defects activated
by testers using manual approach and Tapir framework}
\end{figure*}

Figure \ref{fig:Unique-inserted-defects} provides details on another
comparison of the unique inserted defects that were activated during
the activity of individual testers in both groups.

\subsection{The Results of Case Study 2}

Table \ref{tab:Comparison-navigational-strategies} presents a comparison
of different Tapir framework navigational strategies (refer to Table
\ref{tab:Navigational-strategies}) based on the data that were automatically
collected from the SUT model. In Table \ref{tab:Comparison-navigational-strategies},
we use the metrics that were previously defined in Table \ref{tab:Metrics-used-to}.
In this case study, we excluded steps longer than 15 minutes and assumed
that such length was caused by leaving the session open and not testing. 

In Group 1, the excluded steps represented 0.54\%, 0.87\%, 0.49\%
and 0.76\% of the total recorded steps in Groups 2, 3, and 4, respectively.

\begin{table*}
\caption{\label{tab:Comparison-navigational-strategies} Comparison of Tapir
navigational strategies based data from SUT model}
\centering{}%
\begin{tabular}{|>{\centering}p{5cm}|>{\centering}p{2.8cm}|>{\centering}p{3cm}|>{\centering}p{2.8cm}|>{\centering}p{2.8cm}|}
\hline 
Metric & Group 1 & Group 2 & Group 3 & Group 4\tabularnewline
\hline 
\hline 
Navigational strategy & RANK\_NEW\_ TEAM & RANK\_NEW & RANK\_NEW \_TEAM & RANK\_NEW\tabularnewline
\hline 
Ranking function & \textit{ElementTypeRank} & \textit{ElementTypeRank} & \textit{PageComplexityRank} & \textit{PageComplexityRank}\tabularnewline
\hline 
Test data strategy & DATA\_NEW\_ RANDOM\_TEAM & DATA\_NEW\_ RANDOM & DATA\_NEW\_ RANDOM\_TEAM & DATA\_NEW\_ RANDOM\tabularnewline
\hline 
\hline 
$\mid T\mid$ & 13 & 11 & 12 & 12\tabularnewline
\hline 
\textit{pages} & 224.0 & 211.7 & 233.6  & 206.2\tabularnewline
\hline 
\textit{u\_pages} & 47.1 & 39.0 & 51.4 & 42.2\tabularnewline
\hline 
\textit{r\_pages} & 21.0\% & 18.4\% & 22.0\% & 20.5\%\tabularnewline
\hline 
\textit{links} & 131.8 & 104.2 & 142.5 & 118.0\tabularnewline
\hline 
\textit{u\_links} & 54.4 & 37.9 & 58.1 & 41.1\tabularnewline
\hline 
\textit{r\_links} & 41.3\% & 36.4\% & 40.8\% & 34.8\%\tabularnewline
\hline 
\textit{actions} & 69.3 & 57.5 & 75.1 & 62.7\tabularnewline
\hline 
\textit{u\_actions} & 34.8 & 24.6 & 38.3 & 29.6 \tabularnewline
\hline 
\textit{r\_actions} & 50.2\% & 42.8\% & 51.0\% & 47.2\%\tabularnewline
\hline 
\hline 
\textit{time\_page} & 17.8 & 19.1 & 19.0 & 21.4\tabularnewline
\hline 
\textit{time\_u\_page} & 84.6 & 103.6 & 86.3 & 104.7\tabularnewline
\hline 
\textit{time\_link} & 30.2 & 38.8 & 31.1 & 37.4\tabularnewline
\hline 
\textit{time\_u\_link} & 73.2 & 106.6 & 76.3 & 107.5\tabularnewline
\hline 
\textit{time\_action} & 57.5 & 70.3 & 59.0 & 70.4\tabularnewline
\hline 
\textit{time\_u\_action} & 114.5 & 164.3 & 115.8 & 149.2\tabularnewline
\hline 
\hline 
\textit{defects} & 19.7  & 16.8 & 20.3  & 17.3 \tabularnewline
\hline 
\textit{u\_defects} & 8.6 & 6.9 & 9.1 & 7.4\tabularnewline
\hline 
\textit{time\_defect} & 202.2 & 240.6 & 218.4 & 255.3\tabularnewline
\hline 
\textit{time\_u\_defect} & 463.3 & 585.8 & 487.2 & 596.9\tabularnewline
\hline 
\end{tabular}
\end{table*}

The relative differences between the results of the Case Study 2 groups
are presented in Table \ref{tab:REL-DIF-Comparison-navigational-strategies}.

\begin{sidewaystable*}
\caption{\label{tab:REL-DIF-Comparison-navigational-strategies}Relative differences
between results of Case Study 2 groups}
\centering{}%
\begin{tabular}{|>{\centering}p{6.5cm}|>{\centering}p{3.3cm}|>{\centering}p{3.3cm}|>{\centering}p{3.3cm}|>{\centering}p{3.3cm}|}
\hline 
Relative difference formula ($\alpha$ stands for a metric from Table
\ref{tab:Comparison-navigational-strategies}) & $\frac{\alpha_{Group1}-\alpha_{Group2}}{\alpha_{Group1}}\cdot100\%$ & $\frac{\alpha_{Group3}-\alpha_{Group4}}{\alpha_{Group4}}\cdot100\%$ & $\frac{\alpha_{Group2}-\alpha_{Group4}}{\alpha_{Group2}}\cdot100\%$ & $\frac{\alpha_{Group1}-\alpha_{Group3}}{\alpha_{Group1}}\cdot100\%$\tabularnewline
\hline 
Metric / Comment & RANK\_NEW vs. RANK\_NEW\_TEAM for \textit{ElementTypeRank}  & RANK\_NEW vs. RANK\_NEW\_TEAM for \textit{PageComplexityRank} & \textit{ElementTypeRank} vs. \textit{PageComplexityRank} for RANK\_NEW  & \textit{ElementTypeRank} vs. \textit{PageComplexityRank} for RANK\_NEW\_TEAM \tabularnewline
\hline 
\hline 
\textit{pages} & 5.5\% & 11.7\% & 2.6\%  & -4.1\% \tabularnewline
\hline 
\textit{u\_pages} & 17.2\% & 17.9\% & -8.2\% & -8.4\%\tabularnewline
\hline 
\textit{r\_pages} & 12.4\% & 7.0\% & -11.1\% & -4.4\%\tabularnewline
\hline 
\textit{links} & 20.9\% & 17.2\% & -13.2\% & -7.5\%\tabularnewline
\hline 
\textit{u\_links} & 30.3\% & 29.3\% & -8.4\% & -6.4\%\tabularnewline
\hline 
\textit{r\_links} & 11.9\% & 14.6\% & 4.2\% & 1.2\%\tabularnewline
\hline 
\textit{actions} & 17.0\% & 16.5\% & -9.0\% & -7.7\%\tabularnewline
\hline 
\textit{u\_actions} & 29.3\% & 22.7\% & -20.3\% & -9.1\%\tabularnewline
\hline 
\textit{r\_actions} & 14.8\% & 7.4\% & -10.3\% & -1.5\%\tabularnewline
\hline 
\hline 
\textit{time\_page} & -7.4\% & -12.9\% & -12.2\%  & -6.3\%\tabularnewline
\hline 
\textit{time\_u\_page} & -22.5\% & -21.3\% & -1.0\% & -1.9\%\tabularnewline
\hline 
\textit{time\_link} & -28.3\% & -20.3\% & 3.5\% & -2.8\%\tabularnewline
\hline 
\textit{time\_u\_link} & -45.6\% & -40.8\% & -0.8\% & -4.0\%\tabularnewline
\hline 
\textit{time\_action} & -22.3\% & -19.3\% & -0.2\% & -2.6\%\tabularnewline
\hline 
\textit{time\_u\_action} & -43.5\% & -28.9\% & 9.2\% & -1.1\%\tabularnewline
\hline 
\hline 
\textit{defects} & 14.7\%  & 14.8\% & -3.0\% & -3.0\%\tabularnewline
\hline 
\textit{u\_defects} & 19.8\% & 18.7\% & -7.2\% & -5.5\%\tabularnewline
\hline 
\textit{time\_defect} & -19.0\% & -16.9\% & -6.1\% & -7.4\%\tabularnewline
\hline 
\textit{time\_u\_defect} & -26.5\% & -22.5\% & -1.9\% & -4.9\%\tabularnewline
\hline 
\end{tabular}
\end{sidewaystable*}

\subsection{\label{subsec:Results-of-the-case-study-3}The Results of Case Study
3}

Table \ref{tab:Comparison-TAPIR-vs-MANUAL-data-Case Study 3} presents
a comparison of the manual exploratory testing approach with the Tapir
framework for the experiment with the Pluto system. Compared with
Case Study 1, the real defects were present in the SUT code in this
case study (refer to Section \ref{subsec:System-Under-Test}). In
the Tapir framework, the RANK\_NEW navigational strategy with ranking
functions \textit{ElementTypeRank} and \textit{PageComplexityRank}
were employed. DATA\_NEW\_RANDOM was utilized as the test data strategy.

The data collection method was the same as the data collection method
in Case Study 1, including the meaning of $\mathit{DIFF}$ in Table
\ref{tab:Comparison-TAPIR-vs-MANUAL-data-Case Study 3}. To evaluate
the experiment, the metrics defined in Table \ref{tab:Metrics-used-to}
are employed. In the statistics, we excluded test steps longer than
15 minutes on a particular page. We considered the possibility that
the session was opened but testing was not performed. In the case
of the manual exploratory testing, these excluded steps represented
1.26\% of the total recorded steps, and in the case of Tapir framework
support, this ratio was 0.64\%.

\begin{table*}
\caption{\label{tab:Comparison-TAPIR-vs-MANUAL-data-Case Study 3} Comparison
of manual exploratory testing approach with Tapir framework: data
from SUT model for Case Study 3}
\centering{}%
\begin{tabular}{|c|c|c|c|}
\hline 
Metric & Manual approach & Tapir framework used & $\mathit{DIFF}$\tabularnewline
\hline 
\hline 
$\mid T\mid$ & 10 & 10 & -\tabularnewline
\hline 
\textit{pages} & 98.6  & 122.5  & 19.5\%\tabularnewline
\hline 
\textit{u\_pages} & 24.3 & 33.1 & 26.6\%\tabularnewline
\hline 
\textit{r\_pages} & 24.6\% & 27.0\% & 8.8\%\tabularnewline
\hline 
\textit{links} & 39.3 & 61.5 & 36.1\%\tabularnewline
\hline 
\textit{u\_links} & 16.1 & 26.9 & 40.1\%\tabularnewline
\hline 
\textit{r\_links} & 41.0\% & 43.7\% & 6.3\%\tabularnewline
\hline 
\textit{actions} & 13.6 & 25.5 & 46.7\%\tabularnewline
\hline 
\textit{u\_actions} & 8.2 & 17.7 & 53.7\%\tabularnewline
\hline 
\textit{r\_actions} & 60.3\% & 69.4\% & 13.1\%\tabularnewline
\hline 
\hline 
\textit{time\_page} & 23.9 & 22.3  & -7.4\% \tabularnewline
\hline 
\textit{time\_u\_page} & 97.1 & 82.5 & -17.8\%\tabularnewline
\hline 
\textit{time\_link} & 60.1 & 44.4 & -35.3\%\tabularnewline
\hline 
\textit{time\_u\_link} & 146.6 & 101.5 & -44.5\%\tabularnewline
\hline 
\textit{time\_action} & 173.5 & 107.0 & -62.1\%\tabularnewline
\hline 
\textit{time\_u\_action} & 287.8 & 154.2 & -86.6\%\tabularnewline
\hline 
\hline 
\textit{defects} & 34.8 & 44.1 & 21.1\%\tabularnewline
\hline 
\textit{u\_defects} & 22,2 & 28.7 & 22.6\%\tabularnewline
\hline 
\textit{time\_defect} & 67,8 & 61.9 & -9.6\%\tabularnewline
\hline 
\textit{time\_u\_defect} & 106.3 & 95.1 & -11.8\%\tabularnewline
\hline 
\end{tabular}
\end{table*}

\begin{figure*}
\centering{}\includegraphics[width=18cm]{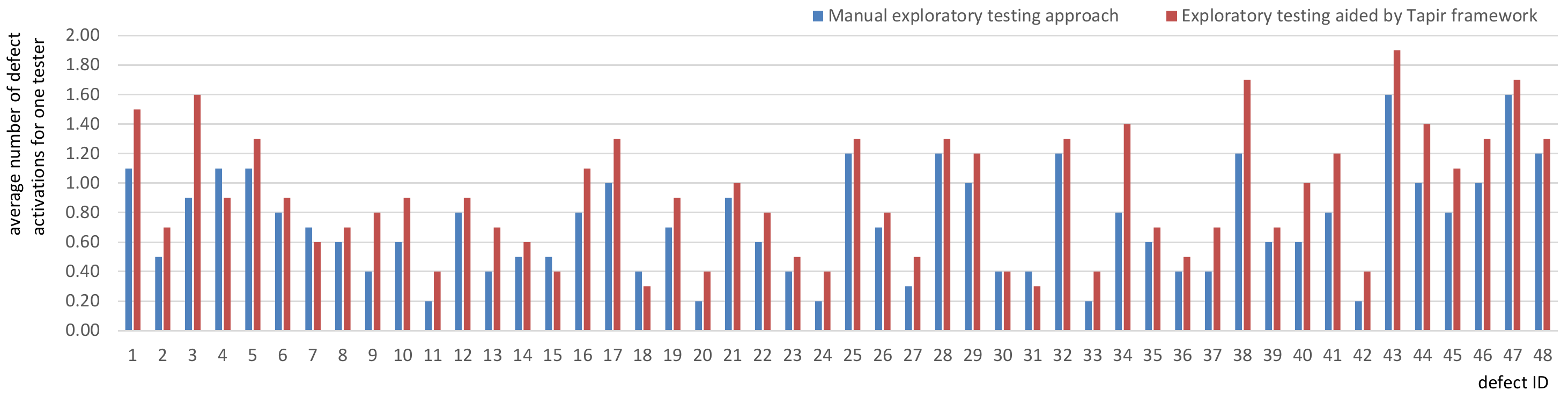}\caption{\label{fig:Potential-to-detect-defects-Case Study 3}Potential of
manual exploratory testing and Tapir framework approach to detect
injected defects in the SUT in Case Study 3}
\end{figure*}

To measure the activated defects, we logged the flow calls of the
SUT method during the tests by the added logging mechanism. Then,
we automatically compared these logs with the recorded SUT model to
determine which defects were activated during certain test steps.
In this case study, all defects were activated by both groups. The
average number of defects activated by both groups are depicted in
Figure \ref{fig:Potential-to-detect-defects-Case Study 3}. 

In this figure, we depict the average number of times one tester activated
the defect (e.g., the value 0.5 indicates that 50\% of the testers
in the group has activated the defect once, and the value 2 indicates
that all testers in the group activated the defect twice).

\begin{figure*}
\begin{centering}
\includegraphics[width=18cm]{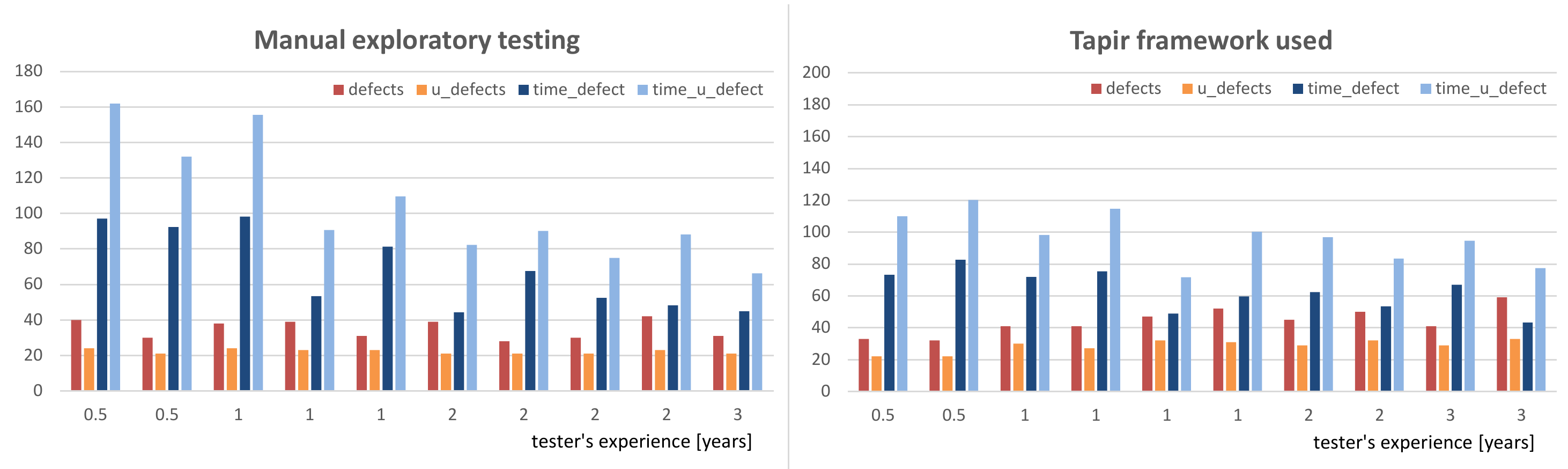}
\par\end{centering}
\caption{\label{fig:Defect-activations-related}Defect activations related
to experience of the testers (Case Study 3)}
\end{figure*}

In Figure \ref{fig:Defect-activations-related} we present defect
activation data related to the experience of the testers. In the graphs,
individual groups of columns present the data for individual testers.
On the x-axis, the praxis of the tester in years is captured. In the
graph, we present \textit{defects}, \textit{u\_defects}, \textit{time\_defec}t
and \textit{time\_u\_defect} values.

\section{\label{subsec:Evaluation-of-the}Discussion}

To evaluate Case Studies 1 and 3, we analyze the data in Table \ref{tab:Comparison-TAPIR-vs-MANUAL-data}
and \ref{tab:Comparison-TAPIR-vs-MANUAL-data-Case Study 3}. In the
case of the MantisBT system, which is the subject of Case Study 1
(Table \ref{tab:Comparison-TAPIR-vs-MANUAL-data}), we note that using
the Tapir framework causes the testers to explore larger extents of
the SUT compared with the manually performed exploratory testing.
This effect can be observed for the total and unique SUT pages (values
\textit{pages} and \textit{u\_pages}), where Tapir support leads the
testers to explore 23.3\% more pages and 41\% additional unique SUT
pages. For the total link elements and unique link elements (\textit{links}
and \textit{u\_links}, respectively) and the total and unique action
elements of the pages (\textit{actions} and \textit{u\_actions}, respectively),
the differences in the values are even higher.

For the Pluto system, which is the subject of Case Study 3 (Table
\ref{tab:Comparison-TAPIR-vs-MANUAL-data-Case Study 3}), the Tapir
framework support increased the value \textit{pages} by 19.5\% and
the value \textit{u\_pages }by 26.6\% as measured by the relative
difference $\mathit{DIFF}$. The number of the total and unique link
and action elements (values \textit{links}, \textit{u\_links}, \textit{actions}
and \textit{u\_actions}) is higher. This case study confirmed the
trend observed in Case Study 1.

However, individual times spent by the exploratory testing process
differ; thus, the efficiency of the exploratory testing process aided
by the framework must be examined in more detail to analyze the proper
relationships of the data. Three key indicators are analyzed: (1)
the ratio of repetition of the pages and page elements during the
testing process, (2) the extent of the SUT explored per time unit,
and (3) the defect detection potential.

\subsection{Repetition of the Pages and Page Elements}

The ratio of the repetition of the pages and ratio of the repetition
of the page elements during the testing process (RQ1, RQ2) indicate
the extent of possible unnecessary action in the SUT during the exploratory
testing process. In the collected data, we express this metric as
the ratio of the unique pages or the elements exercised during the
tests. We start analyzing the data of Case Study 1 (MantisBT, Table
\ref{tab:Comparison-TAPIR-vs-MANUAL-data}). For the Tapir framework,
the ratio of unique pages explored in the exploratory testing process
(value \textit{r\_pages}) is improved by 4.4\% (23.1\% in the relative
difference $\mathit{DIFF}$), the ratio of unique link elements (value
\textit{r\_links}) is improved by 5.8\% (14.8\% in the relative difference
$\mathit{DIFF}$), and the ratio of unique action elements (value
\textit{r\_actions}) presents the largest improvement of 8.8\% (18.8\%
in the relative difference $\mathit{DIFF}$). These improvements are
significant; however, a more detailed explanation is needed to discuss
the relevance of these metrics.

The data indicate that the ratio of unique pages is relatively low.
For example, when exploring the SUT in the manual exploratory testing
process, each page was repeated an average of 6.83 times to achieve
a new page in the SUT. In the case of the Tapir framework support,
this number decreases to 5.25 because the visited pages in the SUT
are repeated during the testing process. An interesting point is that
the links are frequently repeated during the testing process. For
the manually performed exploratory testing, participants exercised
each link 3.02 times to explore one new unique link transition. In
the case of the Tapir framework support, this ratio decreased to 2.57.
When imagining navigation in the SUT and repetition of its particular
functions with various test data, this finding is consistent with
the total picture. The same case is the repetition of action elements,
in which each action has been repeated 2.56 times in the case of the
manual exploratory testing process and 2.08 times in the case of Tapir
support. 

In Case Study 3 (Pluto, Table  \ref{tab:Comparison-TAPIR-vs-MANUAL-data-Case Study 3})
, we observe a similar trend in which the Tapir framework improved
the ratio of unique pages that were explored in the exploratory testing
process (value \textit{r\_pages}) by 2.4\% (8.8\% in the relative
difference $\mathit{DIFF}$) and the ratio of unique action elements
(value \textit{r\_links}), the improvement was 2.8\% (6.3\% in relative
difference $\mathit{DIFF}$) in case of Tapir framework support. In
case of action elements (value \textit{r\_actions}) by 9.1\% (13.1\%
in relative difference $\mathit{DIFF}$).

However, we concluded that the efficiency of the Tapir framework or
exploratory testing process cannot be evaluated solely based on the
ratio of unique elements because exercising the SUT with additional
combinations of test data may decrease these numbers and impact the
efficiency of the testing process. The use of additional test data
combinations (with the SUT elements repaired additional times during
the exploration) can lead to the detection of additional defects.
This outcome strongly depends on the testing goals and principal types
of defects that we want to detect. If the testing goal is a rapid
smoke test of the SUT, the ratio of unique pages or page elements
can be a suitable indicator of process efficiency. If the testing
goal is to detect additional complex structural defects in the SUT,
this metric is not a reliable indicator of the testing efficiency.
Thus, other indicators must be analyzed and discussed.

\subsection{The Extent of the SUT Explored }

The extent of the SUT explored per time unit (RQ1, RQ2) indicates
total time efficiency when exercising the SUT with exploratory tests.
In Case Study 1 (Mantis BT, Table \ref{tab:Comparison-TAPIR-vs-MANUAL-data}),
the average time spent on a page (\textit{time\_page}) improves by
6.6\% in the case of Tapir support. This finding can be explained
by the Tapir handling overhead connected to the exploratory testing
process, including the documentation of the path, decision, and documentation
of the test data. Because the SUT pages were frequently repeated,
the significance of this result is not major. The detailed data in
Figure \ref{fig:Average-times-spent} indicate differences among the
individual times spent on a page by the testers. This factor is strongly
influenced by an individual tester\textquoteright s attitude and work
efficiency.

For the individual time spent on SUT pages, we need to distinguish
two factors that contribute to the total testing time: (1) overhead
related to the exploratory testing process, which is decreased by
the Tapir framework; and (2) time required to analyze the SUT page
and identify and report defects. The second part is equivalent in
both the manual exploratory testing and aided exploratory testing.
In the first factor, the machine support can reduce the time spent
on overhead activities. In the provided data, both parts are mixed
(because distinguishing these two parts is nearly impossible when
collecting data based on monitoring the events in the SUT user front
end).

Because we are interested in exploring the SUT functions available
on pages, we analyze the amount of time is needed to explore the SUT
action element (value \textit{time\_action}) or link (value \textit{time\_link}).
In the case of action elements, the time significantly changes by
97.3\% in the case of Tapir support. In the case of links, the difference
is also significant (43.2\%). The differences are even more striking
in the case of unique pages (value \textit{time\_u\_page}, difference
38.7\% in favor of the Tapir framework), unique action elements (value
\textit{time\_u\_action}, difference 141.9\%), and unique link elements
(value \textit{time\_u\_link}, difference 68.1\%).

In Case Study 3 (Pluto system, Table  \ref{tab:Comparison-TAPIR-vs-MANUAL-data-Case Study 3}),
the Tapir framework improved the time spent on a page (\textit{time\_page})
by 7.4\%. However, an analysis of the data related to the link and
action elements (functions available on SUT pages) provides more relevant
results. With the framework support, the value \textit{time\_action
}increased by 62.1\% and the value \textit{time\_link} increased by
35.3\%. Similar to Case Study 1, this improvement is greater in the
case of unique action elements (value \textit{time\_u\_action}, difference
86.6\%) and unique link elements (value \textit{time\_u\_link}, difference
44.5\%).

From the presented figures, we can conclude that the Tapir framework
leads to a more efficient exploration of the SUT functions in relation
to the time spent testing. However, this optimism can diminish when
we discuss the possible various goals of the testing process. For
a rapid smoke or exploratory lightweight testing of the SUT, when
the primary mission statement is to explore the new SUT parts rapidly
and efficiently, Tapir can provide promising support. For more thorough
testing, the validity of these metrics shall be revised as more thorough
tests and more extensive variants of test data are employed. Thus,
the results for this part will be analyzed based on the efficiency
of the defect detection potential.

\subsection{Defect Detection}

Defect detection potential (RQ1, RQ2)\textbf{ }is an alternative to
the defect detection rate. In this metric, we determine whether the
artificial defect has been activated in the code (which was ensured
by the Tapir framework logging mechanism). When a tester activates
a defect, he is capable of subsequently identifying and reporting
the defect. In Case Study 1 (Mantis BT, Table \ref{tab:Comparison-TAPIR-vs-MANUAL-data}),
we examine the influence of the Tapir framework on the defect detection
potential in the case of inserted artificial defects.

In the case of Tapir framework support, testers activated a total
of 32.8\% more defects when we considered all activated defects, including
repeating defects (a tester exercised the same functionality with
an inserted defect more frequently) and 31.1\% more defects when we
only considered unique defects. Of the 19 inserted defects, an average
of 4.6 unique defects were detected in the manual execution of exploratory
testing, whereas 6.6 unique defects were detected with Tapir support.
This amount is approximately one-third of all inserted defects, and
this result is attributed to the difficult characteristics of the
inserted artificial defects.

The group that uses the Tapir support exercised the SUT longer than
the group that does not use the Tapir support; therefore, we are interested
in determining the time needed to activate a defect. With Tapir support,
the average time to activate a defect (regardless of whether the activated
defects repeat) was reduced by 21.7\%, whereas for unique defects,
the time was reduced by 18.7\%.

The statistics by individual inserted defects are presented in Figure
\ref{fig:Potential-to-detect-defects}, and the details on the efficiency
of individual participants are provided in Figure \ref{fig:Unique-inserted-defects}.
We observe differences among individual testers concerning their efficiency.
When analyzing the data, we do not observe a direct correlation between
time spent on a page and the number of defects that were detected
by individual testers in both groups.

One of the defects, \textit{mc\_2}, was activated by one tester from
the group that only performed the manual testing and by none of the
testers from the group using the Tapir framework support. This situation
deserves an analysis. The defect \textit{mc\_2} can be activated via
the issue list form by the following sequence: (1) set a combo box
value to determine which operation has to be performed with a selected
list of issues to the \textquotedblleft set/unset sticky\textquotedblright{}
value, (2) select additional new issues in the list (the defect is
only activated for issues in \textquotedblleft new\textquotedblright{}
state), and (3) submit the form by the \textquotedblleft ok\textquotedblright{}
button. In the combo box that determines whether an operation has
to be performed with a selected list of issues, fourteen different
operations need to be tested. Thus, the exploration sequence needed
to detect the defect that was not directly captured in the SUT model
by a link or action element. The sequence was a combination of particular
data values entered in the form (input elements) and an action element.
Because of the design of the navigational strategies and ranking functions
in the current version of the Tapir framework, this defect has \textquotedblleft escaped\textquotedblright{}
the exploration path of the testers. Conversely, one of the manual
testers has attempted the particular combination needed to activate
this defect.

In Case Study 3 (Pluto, Table  \ref{tab:Comparison-TAPIR-vs-MANUAL-data-Case Study 3}),
in which real defects occur in the SUT code, improvements were achieved
when the Tapir framework was utilized by the testers in the experimental
group. With the framework support, a total of 21.1\% additional defects
were activated by testers. This value shows the possible repetitive
activation of the same defect during the exploratory testing process.
When only considering the unique defects, the Tapir framework caused
the testers to activate 22.6\% additional defects. Because the group
that uses the Tapir framework has spent a longer testing time, we
analyzed the time needed to activate a defect. The Tapir support improved
the average time to activate one defect by 9.6\% (\textit{time\_defect}).
When we only consider unique defect activation, this improvement is
11.8\% (\textit{time\_u\_defect}).

In this case study, the defect density was higher; thus, less time
is needed to activate one defect (\textit{time\_defect}) and one unique
defect (\textit{time\_u\_defect}). Improvements gained by the Tapir
framework are slightly lower than the improvements gained in Case
Study 1; however, improvements have also been achieved in the case
of the Pluto system.

\subsection{Evaluation of Individual Strategies}

In\textbf{ }Case Study 2, we compared the efficiency of individual
strategies provided by the framework\textbf{ }(RQ3). In our analysis
of the data, we refer to Table \ref{tab:REL-DIF-Comparison-navigational-strategies}.
We start with a comparison of the individual testing strategy RANK\_NEW
with the team strategy RANK\_NEW\_TEAM (columns $\frac{\alpha_{Group1}-\alpha_{Group2}}{\alpha_{Group1}}\cdot100\%$
and $\frac{\alpha_{Group3}-\alpha_{Group4}}{\alpha_{Group4}}\cdot100\%$).
The results differ by ranking function (\textit{ElementTypeRank} vs.
\textit{PageComplexityRank}); however, general trends are observed
in the data. The team navigational strategy RANK\_NEW\_TEAM increased
the ratio of unique explored pages (\textit{r\_pages}) by 12.4\% for
\textit{ElementTypeRank} and by 7.0\% for \textit{PageComplexityRank}.
In the case of the ratio of unique link elements (\textit{r\_links}),
the improvement is 11.9\% for \textit{ElementTypeRank} and 14.6\%
for \textit{PageComplexityRank}. A similar trend is observed for unique
action elements (\textit{r\_actions}), where the improvement is 14.8\%
for \textit{ElementTypeRank} and 7.4\% for \textit{PageComplexityRank},
which is employed as a ranking function.

The statistics related to the time efficiency of the testing process
provide more relevant data. The strategy RANK\_NEW\_TEAM performs
well. The average time spent on a unique page (\textit{time\_u\_page})
decreased by 22.5\% for \textit{ElementTypeRank} and 21.3\% for \textit{PageComplexityRank}.
The average time spent on a unique link element (\textit{time\_u\_link})
decreased by 45.6\% for \textit{ElementTypeRank} and 40.8\% for \textit{PageComplexityRank}.
The average time spent on a unique action element (\textit{time\_u\_action})
decreased by 43.5\% for \textit{ElementTypeRank} and 28.9\% for \textit{PageComplexityRank}.
These numbers indicate the favorability of the team strategy.

Regarding the average unique activated defects that were logged, RANK\_NEW\_TEAM
increases the result by 19.8\% for \textit{ElementTypeRank} and 18.7\%
for \textit{PageComplexityRank}. The average time to detect one unique
defect decreases by 26.5\% in the case of \textit{ElementTypeRank}
and by 22.5\% in the case of \textit{PageComplexityRank}. These results
correspond to the explored extent of the SUT functions, which is higher
in the case of the RANK\_NEW\_TEAM navigational strategy.

From these relative differences, \textit{ElementTypeRank} performs
better; and in the case of the team version of the navigational strategy,
greater improvements are observed. This conclusion is not accurate;
thus, to assess the efficiency of \textit{ElementTypeRank} and \textit{PageComplexityRank},
we need to independently analyze the data.

\textit{PageComplexityRank} leads to the exploration of a slightly
larger extent of the SUT (Table \ref{tab:REL-DIF-Comparison-navigational-strategies}).
The ratio of unique pages explored is 11.1\% higher for the RANK\_NEW
navigational strategy and 4.4\% higher for the RANK\_NEW\_TEAM navigational
strategy. The ratio of unique forms explored is 10.3\% higher for
the RANK\_NEW navigational strategy.

Regarding the average times spent on a SUT page, link and action elements,
the only significant difference is the average time spent on a page
(\textit{time\_page}), which is improved by 12.2\% for the \textit{ElementTypeRank}
in the case of the RANK\_NEW navigational strategy and by 6.3\% in
the case of RANK\_NEW\_TEAM navigational strategy. \textit{PageComplexityRank}
leads to the exploration of more complex pages, which requires additional
processing time and more time spent on an SUT page. Because the more
complex pages usually aggregate more unexplored links and action elements,
the extent of the explored SUT parts is higher than the case of\textit{
ElementTypeRank }as previously discussed.

\textit{PageComplexityRank} enables the exploration of more action
elements than \textit{ElementTypeRank} (Table \ref{tab:Comparison-navigational-strategies},
value \textit{u\_actions}) because of the combination of the navigational
strategy algorithm with the \textit{PageComplexityRank} ranking function.
The Tapir framework mechanism scans the following pages and prefers
the more complex pages (usually containing more action elements to
explore). Note that in page $w,$ the ranking function \textit{ElementTypeRank}
is calculated for both link elements $l\in L_{w}$ and action elements
$a\in A_{w}$, whereas the \textit{PageComplexityRank} is only calculated
for link elements $l\in L_{w}$ (refer to Tables \ref{tab:Navigational-strategies}
and \ref{tab:Ranks-used-in-navigational-strategies}).

\textit{PageComplexityRank} also slightly increased the average number
of logged unique activated defects by 7.2\% for the RANK\_NEW strategy
and 5.5\% for the RANK\_NEW\_TEAM strategy. The difference in the
average time to detect these defects was not significant.

From all analyzed data, the combination of navigational strategy and
ranking function RANK\_NEW\_TEAM with \textit{PageComplexityRank}
presents the most efficient number of activated inserted defects and
the extent of the explored SUT. However, when considering the time
efficiency to explore the new SUT functions, RANK\_NEW\_TEAM with
\textit{ElementTypeRank} seems to be a better candidate.

\subsection{Influence of Testers Experience}

The last issue to discuss is the relation between the experience of
the testers and the activated defects (RQ4). As the total testing
time differed for individual testers, no clear trend can be observed
from \textit{defects} and \textit{u\_defects} values in Figure \ref{fig:Defect-activations-related}.
However, when we analyze average time to activate a defect (\textit{time\_defec}t)
and average time to activate unique defect (\textit{time\_u\_defect}),
both of there values decrease with tester's experience in case of
manual exploratory testing. This effect has been observed previously
also in a study by Micallef \textit{et al.} \cite{Micallef2016}.
With the Tapir framework support, we can observe a similar trend,
nevertheless, the decrease of \textit{time\_defec}t and \textit{time\_u\_defect}
values is not so significant as in the case of manual testing. This
is because, for more junior testers (0.5 up to 1 year of experience),
the Tapir framework allowed more efficient exploratory testing and
the values \textit{time\_defec}t and \textit{time\_u\_defect} are
lower. For more senior testers (2 up to 3 years of experience), the
difference in \textit{time\_defec}t and \textit{time\_u\_defect} is
not significant.

\subsection{Summary}

To conclude all three case studies, Case Studies 1 and 3 provided
data to answer RQ1 and RQ2. Compared with the manual approach, the
support of the Tapir framework enables the testers to explore larger
extents of the SUT and parts of the SUT that were previously unreached.
This finding is also documented by the number of SUT pages explored
per time unit, which slightly increases in the case of the SUT pages
in favor of the Tapir framework (6.6\% for MantisBT, 7.4\% for Pluto).
However, this difference becomes significant when we consider unique
pages (38.7\% for MantisBT, 17.8\% for Pluto), total links (43.2\%
for MantisBT, 35.3\% for Pluto), unique links (68.1\% for MantisBT,
44.5\% for Pluto), total action elements (97.3\% for MantisBT, 62.1\%
for Pluto), and unique action elements (141.9\% for MantisBT, 86.6\%
for Pluto). However, these figures only documented the ability of
the framework to enable more efficient exploration of new parts of
the SUT by the testers, and a relationship to the intensity of testing
is not expressed here. More thorough testing involves repetition of
the SUT parts. Thus, the reliability of this indicator will be discussed
in this case. The measured defect detection potential also indicates
the superiority of the Tapir framework. With systematic support (and
because the testers were able to explore a larger extent of the SUT),
the testers activated 31.1\% additional unique inserted defects in
MantisBT and 22.6\% in the Pluto system. Regarding the time efficiency
to detect a defect, in the case of the Tapir framework, this indicator
improved by 18.7\% for the unique inserted defects for MantisBT and
11.8\% for the Pluto system. Case Study 3 differed from Case Study
1 in the defects that were activated during the experiment. Instead
of the artificial defects used in Case Study 1, the SUT code, which
was the subject of Case Study 3, contained a set of real historical
defects. Improvements achieved by the Tapir framework in Case Study
3 were slightly less than the improvements achieved by the Tapir framework
in Case Study 1. However, improvements can also be observed in Case
Study 3. Regarding RQ2, the analysis of the results did not indicate
a decrease in the efficiency of the exploratory testing process supported
by the Tapir framework. 

Regarding the RQ2, no indicator documenting an aspect in which efficiency
of the exploratory testing process supported by the Tapir framework
would decrease was found during the analysis of the results.

Case Study 2 provided data to answer RQ3. Regarding the comparison
between the RANK\_NEW and RANK\_NEW\_TEAM navigational strategies,
the team navigational strategy performs better in all measured aspects.
For the ranking function, the \textit{PageComplexityRank} ratio of
unique explored pages increased by 7.0\%, the ratio of unique link
elements increased by 14.6\%, and the ratio of unique action elements
increased by 7.4\%. The average time spent on a unique page improved
by 21.3\%, the average time spent on a unique link improved by 40.8\%,
and the average time spent on unique action elements improved by 28.9\%.
The total unique activated defects that were logged improved by 18.7\%,
and the average time to detect one unique defect improved by 22.5\%.
Separately analyzed, the ranking function \textit{ElementTypeRank}
in terms of the extent of explored SUT functions and generates a slightly
higher number of activated unique defects.

The combination of the navigational strategy RANK\_NEW\_TEAM with
the ranking functions \textit{ElementTypeRank} and \textit{PageComplexityRank}
does not indicate a preference. RANK\_NEW\_TEAM with \textit{PageComplexityRank}
performed slightly better regarding the extent of explored SUT and
detected defects; conversely, RANK\_NEW\_TEAM with \textit{ElementTypeRank}
was slightly more time efficient.

Regarding the RQ4, average times needed to activate a defect decreased
with tester's experience in case of manual exploratory testing and
slightly in case of Tapir framework support. With the Tapir framework
support, these times decreased significantly for junior testers (0.5
up to 1 year of experience). For more senior testers (2 up to 3 years
of experience), the improvement in these indicators was not significant.

\section{\label{sec:Threats-to-validity}Threats to Validity}

During the case study experiments, we attempted to equalize the conditions
of the compared groups and compare only comparable alternatives while
keeping other conditions fixed for all participant groups. Several
concerns were identified regarding the validity of the data, which
we discuss in this section.

We previously discussed the relevance of the metrics in Section \ref{subsec:Evaluation-of-the},
which was intended to be objective. For each of the key metrics, proper
disclaimers and possible limiting conditions are described. In all
studies, we employed a defect activation concept instead of a defect
detection concept. Defect activation expresses the likelihood that
the tester will notice the defect when executed. In Case Studies 1
and 2, the Tapir framework logging mechanism exactly logged the fact
that a defect was activated. In Case Study 3, we obtained information
about the presence of the defects in the SUT code employed in the
experiment. The SUT code was accompanied by the logging mechanism,
which recorded the flow calls of the SUT methods during the tests.
Then, we performed an automated analysis of the defects that were
activated by the exercised tests. This analysis ensured the accuracy
of the collected data related to defect activation. In practice, we
can expect a lower real defect reporting ratio because certain activated
defects will not be noticed and reported by the testers. Because this
metric was employed in all comparisons, our opinion is that it can
be used for measuring a trend that is alternative to the defect detection
ratio (which can be biased by individual flaws in defect reporting
by experimental team members). The idle time of a tester can influence
the measured times during a session (which is a likely scenario when
analyzing measured data; refer to the graph in Figure \ref{fig:Average-times-spent}).
In the experiments, we did not have a better option for measuring
the time spent during the testing process. Initially, we experimented
with a subjective tester\textquoteright s report of time spent by
individual testing tasks; however, this method proved to be less reliable
than the automated collection of time stamps related to the tester\textquoteright s
actions in the UI, which was employed in the case studies. We attempted
to minimize this problem by excluding excessively lengthy steps (tester
spent more than 15 minutes on a particular page) caused by leaving
the session open and not testing.

We can consider that the influence of these excluded steps is not
significant because the ratio of these steps was lower than 1.26\%
for the manual exploratory testing for all case studies and lower
than 0.87\% for the exploratory testing supported by the Tapir framework
for all case studies.

In Case Study 1, the group that uses the Tapir framework was larger
(31 versus 23 in the group that performed exploratory testing without
support). The size of the group was sufficiently large to mitigate
this risk, and all testers acted individually. Thus, team synergy
did not have a role in this case study, and all analyzed data were
averages for a particular group.

Previous experience in exploratory testing of the experiment participants
and the natural ability of individuals to efficiently perform this
type of testing can differ among the participants. In the experimental
group, none of the testers were expertly specialized in exploratory
testing, which could favor one of the groups. Participants of Case
Studies 1 and 2 were randomly distributed to the groups, which should
mitigate this issue. In the Case study 3, we distributed the testers
to both groups to have the average length of praxis in both groups
equal 1.5.

Regarding the size of the SUTs (Mantis BT tracker and Pluto), their
workflows and screen-flow model are sufficiently extensive to draw
conclusions regarding all defined research questions. The employed
version of MantisBT consists of 202964 lines of code, 938 application
files and 31 underlying database tables. The utilized version of the
Pluto system consists of 56450 lines of code, 427 application files
and 41 underlying database tables.

\section{\label{sec:Related-work}Related Work }

In this section, we analyze three principal areas relevant to the
presented approach: (1) engineering of the SUT model, (2) generation
of tests from the SUT model and (3) exploratory testing technique.
The first two areas are usually closely connected in the MBT approach.
However, in the Tapir framework, we reengineer the SUT model based
on the SUT via a continuous process, which represents a difference
from the standard MBT approach. 

Current web applications have specific elements that differentiate
them from any other software application, and these specific elements
significantly affect the testing of these applications. The contemporary
development styles of UI construction are not only HTML based, and
the user experience is dynamically enhanced using JavaScript, which
increases the difficulty of identifying the HTML elements. The dynamic
nature of the UI hinders the correct identification of the UI elements.
Crawlers, which are tools that systematically visit all pages of a
web application, are often employed to rapidly collect information
about the structure, content, and navigation relationships between
pages and actions of the web application \cite{Tanida:2013,Dallmeier:2013}.
Although crawlers can rapidly inspect an entire application, the sequence
of steps by a crawl may differ from the sequence made by a manual
tester. Some parts of the application are not reachable by the crawler.
In addition, crawlers can address difficulties when user authorization
is required in the SUT. 

Several solutions for reengineering the SUT model from the actual
SUT were discussed in the literature. As an example, we refer to Guitar
\cite{Nguyen2014}, in which web applications are crawled and a state
machine model is created based on the SUT user interface. A mobile
application version of this crawler, MobiGuitar, has been developed
\cite{amalfitano2015mobiguitar}. As an alternative notation to these
state machine models, the Page Flow Graph can be constructed from
the actual web-based SUT \cite{Polpong:2015:7219604}. The graph captures
the relationship among the web pages of the application, and the test
cases are generated by traversing this graph (all sequences of web
pages). The PFG is converted into a syntax model, which consists of
rules, and test cases are generated from this model. The construction
of the PFG is not described in this paper. An extra step is required
to convert the PFG to a syntax model to generate subsequent tests.

Common elements, including UI patterns, are utilized when developers
create the UI of the applications. Examples of this pattern are Login,
Find, and Search. Users with some level of experience know how to
use the UI or how it should be used. As described by Nabuco \textit{et
al.} \cite{Nabuco:2014:MTC:2723218.2723238}, generic test strategies
can be defined to test these patterns. Pattern-based model testing
uses the domain specific language PARADIGM to build the UI test models
based on UI patterns and PARADIGM-based tools to build the test models
and generate and execute the tests. To avoid the test case explosion,
the test cases are filtered based on a specific configuration or are
randomly filtered. An alternative approach is iMPAcT \cite{Morgado2015,MorgadoN12015},
which analyzes UI patterns in mobile applications. These patterns
can be employed in SUT model reengineering.

Numerous approaches can be employed to generate the tests from the
SUT model. Extensive usage of UML as the design modeling language
implies its use for the MBT techniques. A survey to improve the understanding
of UML-based testing techniques was conducted by Shirole \textit{et
al.} \cite{Shirole:2013:UBM:2492248.2492274}, who who focused on
the usage of behavioral specifications diagrams, e.g., sequence, collaboration,
and state-chart and activity diagrams \cite{Jena:2014:6781352}. The
research approaches were classified by the formal specifications,
graph theory, heuristics of the testing process, and direct UML specification
processing \cite{Shirole:2013:UBM:2492248.2492274}. Use-case models
may represent another source for creating functional test cases as
explored by \cite{lipka2015method}. In addition to UML, other modeling
languages, such as SysML \cite{chang2014} or IFML \cite{Frajtak:2015:TIS:2811411.2811556},
mainstream programming languages, finite machine notations, and mathematical
formalisms, such as coq \cite{Paraskevopoulou2015}, are utilized.
The SUT user interface is subject to the model-based generation of
test cases, and integration testing is also under investigation \cite{potuzak2014}.

Open2Test test scripts are automatically generated from software design
documents \cite{Tanno:2015:7273458}. These documents are artifacts
of the design process. A design document, design model and test model
for the integration testing tool TesMa \cite{Tanno:2015:TCA:2819009.2819147}
are extended with information on the structure of the web page (identification
of the HTML elements). When the software specification is changed,
the latest scripts are regenerated from recent test design documents.
This approach reduces the cost of the maintenance of the test scripts.
However, the design documents must be extended with the detailed information
before the test generation process can start.

The code-based API testing tool Randoop\footnote{https://randoop.github.io/randoop}
generates random sequences of method calls for a tested class. These
sequences are stored as JUnit test cases. Klammer \textit{et al.}
\cite{Klammer:2016} have harnessed this tool to generate numerous
random interaction test sequences. The tool cannot interact with the
UI of a tested application; therefore, a reusable set of adapters
that transforms the API calls into UI events was created. The adapters
also set up and tear down the runtime environment and provide access
to the internal state of the SUT. Fraser \textit{et al.} \cite{Fraser:2014:LEA:2702120.2685612}
have evaluated a test case generation tools called EvoSuite on a number
of Java projects (open source, industrial and automatically generated
projects). The large empirical study presented in the paper shows
that EvoSuite can achieve high test coverage but not on all types
of classes (for example the file system I/O classes). The interaction
with the SUT can be recorded using a capture \& replay tool, or the
source code is analyzed to create a call graph, analyzes the event
sequences and later generate the test cases \cite{SanMiguel:2016:GUM:3001854.3001865}.
The event sequences describe the executable actions that can be performed
by the user when employing an application (an Android application,
in this case). The actions are converted to JUnit tests classes that
are executed in Robotium, which is a test automation framework. 

The manual version of the exploratory testing technique has been investigated
by several studies, e.g., \cite{pfahl2014exploratory,Gebizli2017,Micallef2016}.
The influence of a tester\textquoteright s experience on the efficiency
of the testing process has been examined \cite{Micallef2016}. During
this experiment, more experienced testers detect more defects in the
SUT. Teamwork organization in exploratory testing as a potential method
to improve its efficiency was also explored \cite{Raappana2016}.
In this proposal, team sessions are used to organize the work of the
testers. The Tapir framework also employs the idea of teamwork, although
none of the team sessions are organized. However, the framework collects
information about explored parts of the SUT and test data and uses
this information to prevent the duplication of tests within a defined
team of testers.

Conceptually, combinations of exploratory testing and MBT can be traced
in the related literature; however, the main use case of these concepts
differs from the proposed Tapir framework. For instance, recordings
of performed tests are used to detect possible inconsistent parts
of the SUT design model \cite{Gebizli2014}.

In our approach, we use the web application model that was created
via adopting, modifying, and extending the model by Deutsch \textit{et
al.} \cite{deutsch2007specification}. During our study, the model
underwent significant changes, which we explained in Section \ref{subsec:The-SUT-Model}.
From the related study, the model conceptually resembles the IFML
standard. In our previous study, we explored the possibility of using
IFML as an underlying model \cite{Frajtak:2015:TIS:2811411.2811556}.
Specifics of the presented case enable us to retain the model as defined
in this paper. 

\section{\label{sec:Conclusion}Conclusion}

To conduct the exploratory testing process in a resource-efficient
manner, the explored path in the SUT shall be recorded. The entered
test data is remembered to fix defects or perform the regression testing
process, and work is distributed to individual testers in a controlled
and organized manner to prevent duplicate and inefficient tests. This
step can be manually performed; however, this process usually requires
the intensive and permanent attention of a Test Lead. A considerable
amount of time is spent on administrative tasks related to recording
the explored parts of the SUT and other details of tests. A strong
need to communicate the progress of the team of testers and operationally
re-plan the work assignments could render the exploratory testing
technique more challenging for distributed work teams.

For the more extensive SUT, this organization of work can become difficult.
To visualize the current state of the exploratory testing, a shared
dashboard, which is either an electronic dashboard or a physical dashboard
located in an office space, where the exploratory testers are located,
can be utilized. In the case of a more extensive SUT, capturing the
state of testing on a physical dashboard can become problematic, and
the maintenance of data in an electronic dashboard may become a tricky
task. To remember the previously employed test data, shared online
tables can be utilized. However, accessing these tables and maintaining
their content can generate particular overhead for the testers during
the exploratory testing process. Thus, the efficiency gained by a
systematic approach to the test data can be decreased by this overhead.

The proposed Tapir framework aims to automate the administrative overhead
caused by the necessity to document the path in the SUT, test data,
and related information. The framework serves to distribute the work
in the team of exploratory testers by automated navigation based on
several navigational and test data strategies. These strategies are
designed to explore new SUT functions that were previously revealed
by testing, retesting after SUT defect fixes and testing regressions.
To prevent duplicate tests and duplicate test data (which is one of
the risks of the manual exploratory testing process), team versions
of the navigational strategies were designed.

To assess the efficiency of the proposed framework, we conducted three
case studies. In these case studies, we used two systems under testing.
The first system was an open-source MantisBT issue tracker with 19
inserted artificial defects, and it was accompanied by a logging mechanism
to record the activation of the defects during the exploratory testing
process. The second SUT was the Pluto system developed via a commercial
software industrial project with \textcolor{red}{48} real software
defects. The SUT code was accompanied by a logging mechanism that
records the flow calls of the SUT methods during the tests. This mechanism
enabled the automation of exact analyses to determine the defects
that were activated during the tests.

A comparison of the same process performed by the manual exploratory
testing and the exploratory testing supported by the Tapir framework
focuses on the individual exploration of the SUT by particular testers,
and several significant results were obtained. The exploratory testing
supported by the Tapir framework enabled the testers to explore more
SUT functions and to explore components of the SUT that were previously
unreached. This effect was also confirmed by the measured number of
SUT pages explored per time unit, which improved by 6.6\% for MantisBT
and by 7.4\% for the Pluto system in the case of the Tapir framework.
In the case of other elements, this metric significantly improved.
In the case of MantisBT, the improvement was 38.7\% for unique pages,
43.2\% for total links, 68.1\% for unique links, 97.3\% for total
action elements and 141.9\% for unique action elements. In the case
of the Pluto system, the improvement was 17.8\% for unique pages,
35.3\% for total links, 44.5\% for unique links, 62.1\% for total
action elements and 86.6\% for unique action elements.

However, these metrics document the capability of the framework to
enable testers to efficiently explore new parts of the SUT, and they
are the most relevant in the case of rapid lightweight exploratory
testing, in which a tester\textquoteright s goal is to efficiently
explore new SUT functions and as many previously unexplored functions
as possible. In the case of more thorough testing, the relevance of
these metrics will be revised. In principle, SUT pages and elements
would repeat the tests when exercised more frequently by more extensive
input test data combinations. With the support of the Tapir framework,
testers performing exploratory tests of MantisBT were able to reach
and activate 31.1\% more unique inserted artificial defects, and the
time needed to activate one unique defect improved by 18.7\%. In the
experiment with the Pluto system, 22.6\% additional unique real defects
were activated. Regarding the time needed to activate one unique defect,
the improvement was lower relative to MantisBT at 11.8\% but still
significant.

A comparison of navigational strategies and ranking functions clearly
documented that team-based navigational strategies are more efficient
than individual navigational strategies. For the \textit{PageComplexityRank}
ranking function employed in the comparison of the strategies, the
ratio of unique explored pages increased by 7.0\% for the team navigational
strategy. In addition, the ratio of unique link elements increased
by 14.6\%, the ratio of unique action elements increased by 7.4\%,
the average time spent on a unique page improved by 21.3\%, the average
time spent on a unique link improved by 40.8\%, the average time spent
on unique action elements improved by 28.9\%, the total unique activated
defects that were logged improved by 18.7\%, and the average time
to detect one unique defect improved by 22.5\%.

Regarding the ranking functions, \textit{PageComplexityRank} performed
slightly better than \textit{ElementTypeRank} in several aspects.
We are currently conducting additional experiments to obtain an optimal
ranking function. The concept of the Tapir framework enables this
function to be flexibly configured by setting the calibration constants
(refer to \ref{tab:Ranks-used-in-navigational-strategies}). In our
opinion, the structure of the SUT pages and the complexity of its
workflows have a role in determining the optimal ranking function
for a specific case.

Average times needed to activate a defect decreased with tester's
experience. This trend was obvious in case of manual exploratory testing
and also present, however, less significantly, in case of Tapir framework
support. The Tapir framework support decreased times needed to activate
a defect significantly for testers having 0.5 up to 1 year of experience.
For more senior testers, having 2 up to 3 years of experience, no
clear improvement trend was observed.

The results of the manual exploratory testing process can be influenced
by the Test Lead. Even for a manual process, a systematic approach
can be efficient. Providing systematic guidance to the testers (including
proper documentation of the explored SUT parts) is a challenging task
because this activity has to be continuously performed during the
testing process. Thus, the proposed Tapir framework can provide efficient
support in eliminating the administrative overhead and enabling the
Test Lead to focus on the analyzing the state, performing strategic
decisions during the testing process and motivating the testing team.
The effect of this support would be even stronger in the case of extensive
SUTs and exploratory software testing in business domains, with which
the testers do not have familiarity. The results from the presented
case studi\textcolor{black}{es documented the viability of the proposed
concept and motivate us to evolve the framework.}

\section*{Acknowledgments}

This research is conducted as part of the project TACR TH02010296
Quality Assurance System for Internet of Things Technology. The research
is supported by the internal grant of CTU in Prague, SGS17/097/OHK3/1T/13.

\bibliographystyle{IEEEtran}
\bibliography{mybibfile}

\end{document}